\documentclass[journal=tches,preprint]{iacrtrans}

\usepackage{lipsum} 
\usepackage{xspace}
\usepackage[nolist]{acronym}
\usepackage{booktabs}
\usepackage{siunitx}  
\usepackage{standalone}
\usepackage{tikz}
\usepackage{pgfplots}
\usepackage{caption}
\usepackage{subcaption}
\usepackage{xcolor}

\usepackage{orcidlink}

\definecolor{green1}{RGB}{41, 75, 41}
\definecolor{green2}{RGB}{120, 148, 97}

\author{Fabian Schwarz\inst{1}\fnmsep{}\textsuperscript{\orcidlink{0000-0002-8549-3881}}\fnmsep{}\footnote{These authors contributed equally.} \and Jan Philipp Thoma\inst{2}\fnmsep{}\textsuperscript{\orcidlink{0000-0003-1613-732X}}\fnmsep{}\protect\textsuperscript{\thefootnote} \and \\ Christian Rossow\inst{1} \and Tim G\"uneysu\inst{2,3}\fnmsep{}\textsuperscript{\orcidlink{0000-0002-3293-4989}}}
\institute{
  CISPA Helmholtz Center for Information Security, Saarbr\"ucken, Germany, \email{{fabian.schwarz,rossow}@cispa.de}
  \and
  Ruhr University Bochum, Bochum, Germany, \email{first.last@rub.de}
  \and
  DFKI GmbH
}

\newcommand{\toolName}[0]{KeyVisor}
\title[KeyVisor -- An ISA Extension for Protected Key Handles]{\toolName{} -- A Lightweight ISA Extension for Protected Key Handles with CPU-enforced Usage Policies}

\newcommand{\kl}{\toolName\xspace}
\newcommand{\handleCacheL}[0]{Handle State Cache}
\newcommand{\handleCacheS}[0]{HSC}
\newcommand{\dIV}[0]{$IV_{data}$}
\newcommand{\hIV}[0]{$IV_{handle}$} 

\newcommand{\handleHWunit}[0]{handle wrapper}

\newcommand{\remoteKeyImporter}[0]{Remote Key Provisioner}

\newcommand{\handleGenStruct}[0]{handlegen}
\newcommand{\encdecStruct}[0]{I/O structure}

\newcommand{\innerWrapKey}[0]{visor key}
\newcommand{\innerWrapKeySym}[0]{$k_{visor}$}
\newcommand{\userKey}[0]{user key}
\newcommand{\userKeySym}[0]{$k_{user}$}
\newcommand{\keyHandle}[0]{key handle}
\newcommand{\KeyHandle}[0]{Key handle}

\newcommand{\smallsubsec}[1]{\noindent\textbf{#1}}

\ifdefined\algorithmautorefname
\renewcommand{\algorithmautorefname}{Algorithm}
\fi 
\ifdefined\definitionautorefname
\renewcommand{\definitionautorefname}{Definition}
\fi

\begin{document}

\maketitle


\begin{acronym}
 \setlength{\itemsep}{0.2em}
 \acro{AAD}{additional authenticated data}
 \acro{ASID}{address space identifier} 
 \acro{CPU}{central processing unit} 
 \acro{DoS}{Denial of Service}
 \acro{FF}{flip-flop}
 \acro{GCM}{Galois Counter Mode}
 \acused{GCM}
 \acro{HSM}{hardware security module}
 \acrodefplural{HSM}[HSMs]{hardware security modules}
 \acro{ISA}{Instruction Set Architecture}
 \acro{LFSR}{linear feedback shift register}
 \acro{LUT}{lookup table}
 \acro{PMP}{physical memory protection}
 \acro{TEE}{trusted execution environment}
 \acro{TLB}{Translation Lookaside Buffer}
 \acro{TPM}{Trusted Platform Module}
 \acro{TRNG}{true random number generator}
\end{acronym}

\acused{CPU}

\acused{CPU}
\begin{abstract}
The confidentiality of cryptographic keys is essential for the security of protection schemes used for communication, file encryption, and outsourced computation. Beyond cryptanalytic attacks, adversaries can steal keys from memory via software exploits or side channels, enabling them to, e.g., tamper with secrets or impersonate key owners.
Therefore, existing defenses protect keys in dedicated devices or isolated memory, or store them only in encrypted form.
However, these designs often provide unfavorable tradeoffs, sacrificing performance, fine-grained access control, or deployability.

In this paper, we present \toolName{}, a lightweight ISA extension that securely offloads the handling of cryptographic keys to the CPU.
\toolName{} provides CPU instructions that enable applications to request protected \keyHandle{}s and perform AEAD cipher operations on them.
The underlying keys are accessible only by \toolName{}, and thus never leak to memory.
\toolName{}'s direct CPU integration enables fast crypto operations and
hardware-enforced key usage restrictions, e.g., keys usable only for de-/encryption, with a limited lifetime, or with a process binding.
Furthermore, privileged software, e.g., the monitor firmware of TEEs, can revoke keys or bind them to a specific process/TEE.
We implement \toolName{} for RISC-V based on Rocket Chip, evaluate its performance, and demonstrate real-world use cases, including key-value databases, automotive feature licensing, and a read-only network middlebox.
\end{abstract}


\section{Introduction}
Cryptographic operations like data encryption or authentication form the foundation of various security applications, ranging from TLS communication, VPNs, and file protection to device attestation and distributed IoT networks.
In any of the underlying schemes, the secret keys are a critical asset that must be protected against attacks.
If the keys are leaked, all higher-level security mechanisms based on them lose all (or most of) their protection guarantees.
However, as applications require plaintext access to these keys, attackers gaining access to the system memory, for instance, via local software exploitation (e.g., ROP, use-after-free), side-channel attacks, or remote exploits like Heartbleed~\cite{heartbleed14}, can directly leak the keys and thus break the security of the applications.

To mitigate the risk of a key leakage, solutions for secure key management prevent direct key access by applications.
Instead, they protect keys and their cryptographic operations in isolated environments, and provide a controlled interface to the keys via so-called \emph{\keyHandle{}s}.
Applications can request crypto operations via these handles without gaining access to the underlying plaintext keys, thus preventing leakage.
Common approaches to implement such key handles can be grouped into three categories.
First, \acp{HSM}  are dedicated crypto devices that manage keys in isolated memory and support key generation and crypto operations, as well as additional access control and authentication features for them.
A widely-used example is the TCG-standardized \ac{TPM}~\cite{tpmTCG}, or comparable vendor-specific chips, e.g., Apple's Secure Enclave~\cite{appleSecureEnc,opentitan,plutonAsTPM24}.
Second, \acp{TEE} support trusted code and data isolation and have therefore been used to implement secure key managers in software, e.g., based on Arm TrustZone~\cite{AndKeymaster,tzkms18} or Intel SGX~\cite{badSGXkms17,sgxeHSM22,sgxKMRA23}.
Third, the Intel Key Locker CPU extension provides~\cite{KeyLocker2020} CPU-protected AES \keyHandle{}s by introducing CPU instructions that generate and use CPU-encrypted handles for AES block operations (no AES modes).

However, existing technologies face limitations regarding their performance, integration, or enforceable key usage
policies, restricting their use cases.
TPM-like devices~\cite{tpmTCG,appleSecureEnc,opentitan,plutonAsTPM24,chakraborty2019simtpm} support many algorithms and key distribution scenarios. 
Their complex design can make application adoption highly non-trivial.
In addition, they typically provide slower performance than CPU accelerators, rendering them unsuited for high throughput scenarios, e.g., 
network communication.
Furthermore, they can only enforce coarse-grained access policies, but no specific and refined rules based on a caller's process identifier.
While virtual software \acp{HSM} or TPMs provide more flexibility and sometimes higher performance, they are often less secure or more dependent on specific platform technologies, e.g., \acp{TEE}~\cite{sgxeHSM22,vtpm06,ftpmTrustZone16}.
Furthermore, \acp{HSM} and key manager services based on \acp{TEE}~\cite{vtpm06,ftpmTrustZone16,tzkms18,AndKeymaster,badSGXkms17,sgxKMRA23} often suffer from high context switching overhead for invocation of a crypto operation, and are usually not supported on embedded systems.
Moreover, various side-channel attacks against \ac{TEE}-based de-/encryption have leaked cryptographic keys despite their isolation~\cite{Chen2020,Goetzfried2017,Bulck2018}.
The Key Locker CPU extension~\cite{KeyLocker2020} provides a further approach in this context, which is limited in its feature set.
In particular, Key Locker can revoke handles only all at once, and the single securely supported usage policy is to make a handle usable only in kernel space.
Key Locker does further not provide hardware support for encryption modes, limiting control on how handles are used, and does not support secure \keyHandle{} sharing across systems.
Moreover, when using Key Locker within \acp{TEE}, the keys are only secure if Key Locker has been configured with a hardware-generated protection key, and the handles are still vulnerable to leakage via side-channel attacks.

To overcome these limitations, we envision a CPU extension for protected \keyHandle{}s that combines a high-performance but lightweight system integration (HW/SW) with strong key usage controls.
Similar to Intel Key Locker, we assume handles usable only via dedicated CPU instructions, enabling the CPU to mediate all operations.
However, we consider a new CPU-integrated policy engine that can enforce \emph{how} and \emph{by whom} each \keyHandle{} can be used, and enables selective \emph{revocation} of key handles.
Leveraging the register-encoded execution context including the privilege level and process/TEE identifiers, the CPU can pinpoint the caller context.
Furthermore, this context can be augmented with a handle-specific state to track a handle's usage controls and lifetime.
With in-CPU support for authenticated encryption ciphers (AEAD), e.g., AES-GCM and ChaCha20-Poly1305, the CPU policy engine can enforce \emph{how} \keyHandle{}s can be cryptographically used, including the ciphers and operations (en-/decrypt) permitted by a caller.
Such fine-grained policies enable security schemes beyond those supported by \acp{TPM} or Key Locker.
By adding support for secure remote key imports, even more become feasible, ranging from read-only TLS inspection to CPU-enforced license keys.

In this paper, we present \emph{\toolName{}}, a CPU extension design for protected \keyHandle{}s that follows the principles outlined above.
\toolName{} provides fast, easy-to-use crypto operations for AEAD ciphers (authenticated encryption with associated data), with user-defined handle restrictions enforced by the CPU.
Like Intel Key Locker, \toolName{} builds on handles that are wrapped by a CPU-internal key and can only be unwrapped by the CPU.
In addition, \toolName{} introduces several CPU-enforced usage restriction policies that can be associated with a \keyHandle{}, and enables fine-grained handle revocation---all without sacrificing AEAD performance.
\toolName{} focuses on AEAD ciphers as they are widely used due to their combination of encryption and authentication, e.g., 
AES-GCM in TLS or ChaCha20-Poly1305 in the Wireguard VPN~\cite{wireguard17}.
Furthermore, AEAD ciphers enable us to design additional en-/decrypt-only key controls, not possible for many non-AEAD (stream) ciphers, e.g., AES-CTR.
In particular, our current design of \toolName{} for RISC-V, based on AES-GCM, allows to control (1)~the permitted AES-GCM operations (encrypt/decrypt), (2)~permitted caller context, e.g., process, (3)~counter-based handle revocation (e.g., to enforce a one-time key), and (4)~selective on-demand handle revocation.
In addition, \toolName{} introduces a local trusted key provisioner that enables remote services to securely export keys as restricted \keyHandle{}s to the local system---without leaking the plaintext key to the local software.

We implement \toolName{} for the RISC-V Rocket Chip CPU and release it as an open-source project for the research community\footnote{\label{fn:proto}\toolName{}'s open-source code will become available on publication of the finished conference paper.}.
\toolName{} introduces four new CPU instructions tailored to creating, using, and revoking \keyHandle{}s, as well as an efficient CPU-internal handle caching structure.
We demonstrate the benefits of \toolName{}'s usage policies by integrating it into three real-world use cases: web services using a vulnerable key-value database, a(n offline) licensing scheme for automotive pay-per-use features, and a read-only network middlebox for guaranteed tamper-free traffic analysis.
Furthermore, we	evaluate the area overhead of \toolName{}'s lightweight design, and measure its performance using micro-benchmarks, comparisons to widely-used crypto libraries, and a TLS-based real-world use case.

In summary, we make the following contributions:
\begin{itemize}
	\item We design \toolName{}, a lightweight ISA extension for protected \keyHandle{}s with CPU-enforced usage policies.
	
	\item We provide a RISC-V hardware implementation of \toolName{} for AES-GCM and evaluate its area requirements and performance, showing its practical feasibility.
	
	\item We demonstrate \toolName{}'s benefits and flexible key usage policies by integrating it in three real-world use cases.
\end{itemize}
\section{Design Goals and Threat model}
\label{kl:sec:motivationProblem}
Existing solutions for protected \keyHandle{}s---most prominently including TPM-like co-processors~\cite{tpmTCG,appleSecureEnc,opentitan,plutonAsTPM24,chakraborty2019simtpm} or the Intel KeyLocker ISA extension~\cite{KeyLocker2020}---are tailored to specific use cases.
TPM-like devices focus on storing and supporting many types of crypto keys and ciphers, measured boot~\cite{tpmboot}, and are limited in high-performance settings.
In contrast, Intel Key Locker achieves high-performance via an in-CPU AES accelerator but has narrow key control, mainly limited to hiding the plaintext AES keys from memory.
Therefore, we envision \toolName{}'s design to enable a wide set of new use cases ranging from high-performance to embedded settings.
We regard a lightweight ISA extension as a key enabler for achieving this, because (1.)~it enables direct access to the CPU register values and thus execution contexts---essential for fine-grained handle controls---and (2.)~provides high-performance AEAD and policy operations while (3.)~being affordable for high-end and low-end CPUs alike.
We will now motivate \toolName{}'s goals (\textit{$\text{G}_{\text{x}}$}) based on concrete use cases where TPM-like solutions and Intel KeyLocker are insufficient, before summarizing the threat model and presenting our full design.

\smallsubsec{UC-1: Preventing leakage via Process Binding.}
Existing \keyHandle{} solutions lack fine-grained control on \emph{who} can use a handle.
While \keyHandle{}s mitigate remote attackers from stealing keys as the handles are unusable on the remote host, handles are still vulnerable to local leakage.
For instance, consider a web service that uses a local key-value database (e.g., Redis) for storing ephemeral web session data.
The web service wants to prevent the database from reading the stored data to protect against a potential compromise of the database.
Therefore, the web service passes values only in encrypted form to the database.
However, if the encryption keys of the web service are leaked, e.g., via a remote information leakage vulnerability or a local side-channel attack, attackers can decrypt the user data stored in the database (e.g., session credentials).
Even if keys are wrapped inside handles, local attackers can exfiltrate and abuse the handles.
The underlying issue is the incapability to enforce fine-grained local controls on what execution contexts are permitted to use a \keyHandle{}.
Therefore, we envision a new \keyHandle{} extension to be able to \textit{bind handles to specific local processes} (\textit{$\text{G}_{\text{1}}$}) or \textit{privilege levels} (\textit{$\text{G}_{\text{2}}$}).
That way, the web service could bind the handles to its own process, guaranteeing 
not only that the keys never leak to remote parties, but also that local attacker processes cannot abuse the \keyHandle{}s.

\smallsubsec{UC-2: Key Revocation and Remote Control.}
Two important concepts for \keyHandle{}s include revocation and remote provisioning.
Revocation enables users to make handles unusable, e.g., for a key rollover or if key usage should be denied after a security incident.
However, TPMs and Intel Key Locker only feature coarse-grained control for set-wise revocation---Key Locker even supporting only an all-or-nothing revocation.
For some use cases, it would be beneficial to even have the CPU automatically revoke a given handle after a specified number of crypto operations, e.g., to enforce one-time handles.
Such a feature is particularly interesting when a remote service wants to temporarily provision a key to a client system, with CPU-enforced lifetime restrictions.
That is, export a key as a restricted handle, without leaking the plaintext key.

As an example, consider an automotive system for which the vendor offers extra features locked behind a licensing system.
Customers can purchase temporary licenses to gain access to specific features for a limited number of activations, e.g., a sport mode providing more motor power.
As part of a feature activation, the car connects to a remote vendor service to check for a valid license.
However, in such a design, customers cannot user their purchased features if the car is in offline mode (e.g., no coverage).
Therefore, vendors require a secure mechanism to locally verify licenses.

One solution could include the remote service provisioning a feature-specific signing key to the car, that enables the user to locally authenticate feature enabling from the respective control unit, e.g., motor unit.
By wrapping the key in a handle, it can be isolated from local system-level attackers trying to bypass the licensing enforcement.
If the number of signing operations could be limited to the number of acquired feature uses, the remote vendor could ensure that the handle is automatically revoked, even if the car is in offline mode.
However, neither TPMs nor Key Locker support such counter-based lifetime policies, and Key Locker does not even support remote key provisioning.
Therefore, for \toolName{}, we envision two new features: (1.)~\textit{selective \keyHandle{} revocation} (\textit{$\text{G}_{\text{3}}$})---on demand, or via lifetime counters---and (2.)~\textit{confidential remote key provisioning (\textit{$\text{G}_{\text{4}}$})}.
That way, the key could be securely provisioned to \toolName{} and transformed into a counter-restricted handle, which is automatically revoked by the CPU when the number of licensed feature activations has been reached.

\smallsubsec{UC-3: Share Read-only Access to Encrypted Data.}
Beyond controlling \emph{by whom} and \emph{how long} \keyHandle{}s can be used, we deem control on \emph{how} they can be used important.
For instance, consider local enterprise clients that communicate with external services via end-to-end-encrypted (E2EE) connections.
The company might want to deploy a remote traffic monitoring service (e.g., on-path intrusion detection/prevention system, short: IDS/IPS) to scan the plaintext traffic for suspicious activities, e.g., data leakage or malware.
However, the company wants to guarantee that the monitor preserves the integrity of the connections to prevent a compromised service from tampering with the traffic.
Common approaches for live traffic decryption either cannot rule out data tampering, e.g., those directly sharing connection keys or performing certificate-based MITM attacks, or require dedicated ports of monitoring services into trusted hardware~\cite{safebricks18,lightbox19}, or protocol changes that enforce read-only access to the plaintext traffic~\cite{lee2019matls}.
Instead, the company wants to wrap the connection keys in \keyHandle{}s that deny encryption operations required for data injections.
However, TPMs do not control if software uses handles for encrypt or decrypt operations, and is too slow for live network monitoring.
Intel Key Locker is CPU-accelerated and can selectively deny de-/encrypt AES block operations.
However, due to Key Locker's lacking in-CPU support for AEAD ciphers, attackers can bypass the de-/encrypt-only handle restrictions, e.g., by \emph{decrypting} TLS traffic using AES-GCM \emph{encrypt} operations (cf.~\autoref{kl:sec:encdecissues} for details).
Therefore, we desire \keyHandle{}s that have full control of the AEAD ciphers in hardware and are adjusted to \textit{securely enforce encrypt-~or decrypt-only handles (\textit{$\text{G}_{\text{5}}$})}.
That way, the company clients could remotely provision decrypt-only handles to the traffic monitoring service, enabling it fast but read-only traffic access, without requiring TLS changes.

\subsection{Threat Model}
\label{kl:subsec:threats}
We now summarize the threat models that we derived for \toolName{} from the above use cases.
We assume that user space or system services (e.g., kernel) use cryptographic keys (here: AES keys) as a basis for their higher-level security protocols.
The services are benign but vulnerable, i.e., they can securely generate and protect their keys on first service startup, but all services might eventually become compromised or target of an attack.
The attacker's goal is to gain uncontrolled access to the plaintext keys in order to abuse the keys for attacks against the security protocols (cf.~use cases).
As the CPU is assumed to be trusted, the services want to protect their keys using CPU-provided \keyHandle{}s to prevent their plaintexts from leaking and to securely enforce usage restrictions on them.

In remote provisioning settings, where remote services want to confidentiality share a key as restricted \keyHandle{} with the local system, we assume that the remote systems are fully trusted, while the local system might already be compromised.
That is, the remote keys and associated restrictions must not be leaked or tampered by local attackers before being transformed into protected \keyHandle{}s.

We exclude physical attackers targeting the CPU and its internal memory, e.g., through fault injection or voltage glitch attacks~\cite{pmpFaultinj20,Trouchkine2021,Buhren2021}.
We refer to orthogonal hardware and software defenses for tackling such strong attackers~\cite{compasec23,Stolz2023}, and instead focus on software-level attackers.
In addition, we exclude \ac{DoS} attacks by strong system-level attackers trying to delete or revoke \keyHandle{}s, as such attackers can anyway shutdown processes or the whole system.
Throughout the paper, we will assume trust in the local OS or monitor software to enable the binding of \keyHandle{}s to a specific process or \ac{TEE} context.
However, the security of the \keyHandle{}s and all remaining usage restrictions stay valid within the stricter threat model assuming user, kernel, and monitor software to eventually become compromised.

\section{Design Concepts}
\label{kl:sec:concepts}
In the following, we introduce the design concepts of \emph{\toolName{}}, our hardware extension for protected \keyHandle{}s with CPU-enforced usage policies.
Without loosing generality, we present the concepts tailored to the RISC-V \ac{ISA} and the AES-GCM AEAD cipher.
Our choice is motivated by the open-source nature of RISC-V and the wide usage of AES-GCM, e.g., in TLS.
Note that \toolName{}'s concepts could be transferred to other \acp{ISA} with minor adaptations and could be extended to additional types of (non-AES) AEAD ciphers, e.g., ChaCha20-Poly1305 (cf.~\autoref{kl:sec:discussion}).

\subsection{\toolName{} in a Nutshell}
\label{kl:sec:nutshell}
\begin{figure}
	\centering
	\centering
	\includegraphics[width=.65\columnwidth]{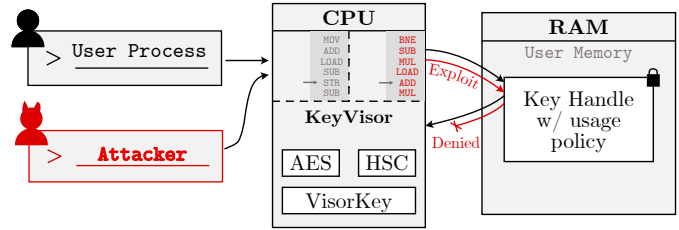}
	\caption{\label{kl:fig:overview}
\toolName{} allows CPUs to replace in-memory keys with protected \keyHandle{}s usable only via new CPU instructions in a policy-defined way, e.g., by the owner process.}
\end{figure}

\toolName{} offloads the confidentiality protection and usage control of AES(-GCM) keys from vulnerable software services to the CPU.
That way, \toolName{} can provide fast cryptographic operations in hardware (de-/encryption) while enforcing strict key isolation based on the CPU-encoded user context.
With \toolName{}, software can use a new CPU instruction to securely wrap their keys (\emph{\userKey{}s}) with a \emph{\innerWrapKey{}} that is only accessible by the \emph{local} CPU (see~\autoref{kl:fig:overview}).
The resulting \keyHandle{}s never leak the plaintext keys and can thus be managed in unprotected memory and storage.
Software can use \keyHandle{}s for cryptographic operations only via new CPU instructions that enforce usage restrictions before securely unwrapping the keys and performing the requested AES-GCM operations efficiently in hardware (e.g., user data encryption).
The plaintext keys can be wiped from memory, thus preventing any leakage to local or remote attackers.

Among \toolName{}'s key concepts that distinguish it from existing key protection solutions are its fine-grained key usage control and revocation management.
When a \userKey{} is transformed into a \keyHandle{}, users can specify a usage restriction policy that governs how and by whom the handle can be used to encrypt or decrypt data.
\toolName{} securely associates the policy with the handle and enforces its rules on each handle-based operation, without sacrificing performance.
The handle policies support high-level rules that specify the permitted AEAD cipher (here: AES-GCM) and types of crypto operations, e.g., ``only permit decryption'', as well as lifetime rules limiting the number of permitted handle uses, e.g., one-time keys.
In addition, \toolName{}'s tight CPU integration---in contrast to external solutions like TPMs---enables context-sensitive policy rules based on the CPU-exposed caller information, e.g., the current process ID or CPU privilege level.
That way, \toolName{} can bind \keyHandle{}s to specific caller contexts, e.g., the kernel or a specific user process, which renders stolen handles unusable even by local attackers.
In addition to restricting handle policies, \toolName{} enables (authorized) software to request revocation of a handle's key via a CPU-internal \keyHandle{} allowlist, which is based on an efficient hardware caching structure, called \emph{\handleCacheL{}} (\emph{\handleCacheS{}}).
Furthermore, \toolName{} introduces an (optional) authenticated \emph{\remoteKeyImporter{}} which can securely receive \userKey{}s and restriction policies from a remote system and forward them to \toolName{}'s handle unit.
That way, remote services can share AES keys that never leak in plaintext to the local software and whose usage is tightly controlled.
In \autoref{kl:sec:usecases}, we will describe how \toolName{} efficiently solves the web, network, and automotive key protection challenges described in \autoref{kl:sec:motivationProblem}, and explore further scenarios in \autoref{kl:sec:discussion}.

\subsection{Transforming Keys into Protected Handles}
\label{kl:sec:genHandles}
\begin{figure}
	\centering
	\includegraphics[width=\linewidth]{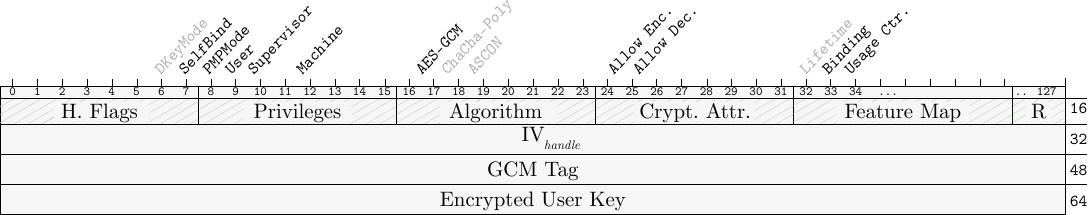}
	\caption{\label{kl:fig:handle_format}The \KeyHandle{} format contains the encrypted \userKey{}, the AES-GCM IV and tag, and the handle usage policy (bits 0 to 128). The policy is signed as GCM AAD.}
\end{figure}

In order to benefit from \toolName{}'s protection guarantees, software services must transform their plaintext AES keys into secure \keyHandle{}s and wipe the plaintext keys from untrusted memory.
\toolName{} adds a new CPU instruction for generating protected \keyHandle{}s that services can directly call.
In \toolName{}, one \keyHandle{} represents one AES \emph{\userKey{}} and its associated usage restriction policies.
In contrast to OS file or socket descriptors which only encode an index into an internal OS table containing all associated data, \toolName{} encodes the key and most of the usage policies directly in the \keyHandle{} object itself.
That way, \toolName{} avoids large expensive CPU-internal memory and lets software store the \keyHandle{}s in untrusted memory or disk storage.

\toolName{} must protect the \keyHandle{}-encoded data against tampering and key extraction.
Otherwise, attackers might overwrite the usage restrictions to gain uncontrolled handle access, or even leak the encoded \userKey{}.
Therefore, \toolName{} introduces a CPU-internal AES-GCM engine and one internal AES key, referred to as the \emph{\innerWrapKey{}}.
\toolName{} uses the AES-GCM engine and \innerWrapKey{} to wrap each \keyHandle{}, i.e., encrypt and sign it.
The \innerWrapKey{} is securely generated by the CPU using a \ac{TRNG} and stored in a protected CPU register only accessible by \toolName{}.
Consequently, only \toolName{} can decrypt the \keyHandle{} and therefore extract and use the \userKey{}.

To be precise, on handle generation, \toolName{} performs an AES-GCM operation in hardware that uses the \innerWrapKey{} \innerWrapKeySym{} to authentically encrypt the user-provided key \userKeySym{}.
A handle usage policy is derived from the user-specified restrictions and is used as \ac{AAD}, i.e., it is signed but not encrypted.
That way, the policy is bound to \userKeySym{}, protected against tampering, and still readable by user software.
For each new \keyHandle{}, \toolName{} adds an entry to its CPU-internal allowlist which keeps track of all valid handles, as we will explain in \autoref{kl:sec:revoke} and \autoref{kl:sec:cpu_mem}.
The resulting \keyHandle{} is shown in \autoref{kl:fig:handle_format}, consisting of the encrypted \userKeySym{}, the authentication tag and initialization vector \hIV{} of the authentic encryption, and the usage policy.
\toolName{} calculates a fresh \hIV{} for each handle using a \ac{LFSR} (cf.~\autoref{kl:sec:concept_encdec} and \autoref{kl:sec:ivGen}) to prevent IV collisions.
Otherwise, IV collisions would break the security of AES-GCM and thus allow for leaking information on the associated \userKey{}s.
We will explain the usage policies of \toolName{}'s handles in \autoref{kl:sec:usagePolicy}.

\subsection{Handle-based Data De-/Encryption}
\label{kl:sec:concept_encdec}

Software services can use \toolName{}'s protected \keyHandle{}s to perform AEAD operations (encrypt-sign, decrypt-verify) with the policy-permitted cipher, e.g., AES-GCM.
\toolName{} ensures that only valid handles can be used and that their restriction policies are securely enforced by the CPU on each operation---without leaking information on the plaintext key.
\toolName{} defines one CPU instruction for encryption and decryption respectively that can be directly called by software in line with the policy, e.g., permitted privilege levels.
The instructions take two input registers storing memory pointers to the key handle and an \encdecStruct{} containing pointers to the required plaintext/ciphertext and cryptographic metadata, e.g., the authentication tag for decryption.

On instruction execution, \toolName{} checks the \keyHandle{} and usage policies before performing the requested crypto operation.
\toolName{} first loads the \keyHandle{} and unwraps it, i.e., verifies its integrity and decrypts the embedded \userKey{} using the \innerWrapKey{} and \ac{GCM} tag.
If an attacker has tampered with the \keyHandle{} data, including \hIV{} and usage policy (used as \ac{AAD}), the unwrapping fails and denies the operation.
Upon success, \toolName{} checks if the handle is valid using its internal allowlist (\handleCacheS{}, \autoref{kl:sec:cpu_mem}) and enforces the handle-associated usage restrictions, e.g., permitted operations or process-binding (discussed in the next section).
If any of the checks fails, the requested crypto operation (encrypt/decrypt) is denied.
By completely loading and checking the handle and its restriction policy before usage, \toolName{} prevents Time-of-Check-to-Time-of-Use attacks~\cite{JWCP2005} that try to concurrently tamper with the handle in memory, e.g., to bypass restrictions.
Finally, \toolName{} loads the crypto and user data (block-wise) from the given input addresses and performs the actual AEAD de-/encrypt operation.

In principle, \toolName{} lets the user control the \dIV{} used for encryption and decryption operations, like existing designs (e.g., OpenSSL, Key Locker).
However, there is one important exception: users \emph{cannot} choose the \dIV{} used by encrypt-only \keyHandle{}s.
Otherwise, attackers could exploit that stream-cipher based AEADs allow to \emph{decrypt} data using the \emph{encrypt} operation, bypassing encrypt-only handle restrictions---as we will explain in \autoref{kl:sec:encdecissues}.
Therefore, \toolName{} uses a hardware full-cycle \ac{LFSR}~\cite{Ward2012} to generate a fresh, collision-free \dIV{} on each operation of encrypt-only handles, preventing such attacks.
In that case, the resulting integrity tag and used \dIV{} are output to the \encdecStruct{}.

\begin{figure*}
	\centering
	\includegraphics[width=.9\textwidth]{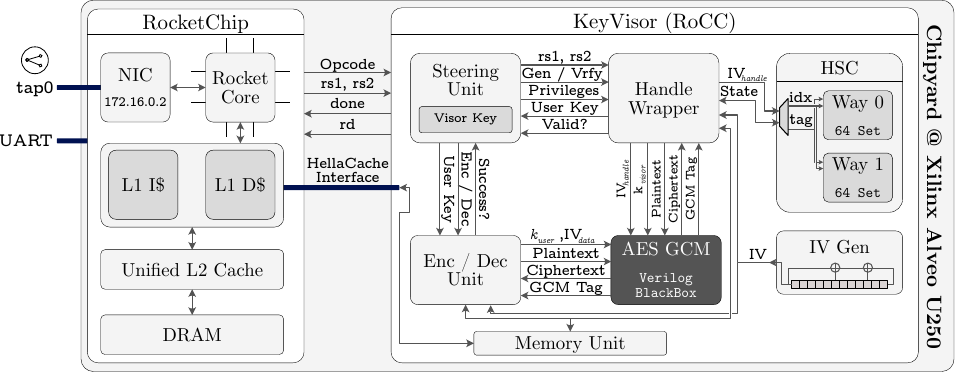}
	\caption{\label{kl:fig:hw_overview}Simplified block diagram of \toolName{}'s hardware implementation.}
\end{figure*}

\subsection{Handle Usage Policies and Revocation}
\label{kl:sec:usagePolicy}
By default, \toolName{}'s \keyHandle{}s would be usable for crypto operations by whoever has memory access to them.
Therefore, to protect handle access, \toolName{} introduces CPU-enforced per-handle usage restriction policies that enable users to easily specify if, how, and by whom \keyHandle{}s can be used.
\toolName{} associates a policy on handle creation (\autoref{kl:sec:genHandles}) and enforces it on each handle-based operation with a minimal overhead.

\smallsubsec{Restrict how handles are used.}
\label{kl:sec:restrictions}
One set of \toolName{}'s restrictions controls \emph{how} \keyHandle{}s can be used. 
This currently includes the selection of the permitted AEAD cipher and cipher operations.
Users can deny encryption or decryption operations for a handle, which enables en-/decrypt-only handles that add asymmetric usage restrictions to otherwise symmetric crypto keys.
In \autoref{kl:sec:nids}, we show how they can enforce read-only access to TLS connections for a traffic monitor.
While \toolName{}'s focus is on AEAD ciphers, in principle, \toolName{} can also support non-AEAD ciphers, e.g., block ciphers like AES-CBC or stream ciphers like AES-CTR.
However, as we will explain in \autoref{kl:sec:encdecissues}, it is not possible to enforce secure en-/decrypt-only handles for existing non-AEAD stream ciphers.

\smallsubsec{Restrict if handles are valid.}
\label{kl:sec:revoke}
The second set of restrictions controls \emph{if} a handle is valid and \emph{when} it becomes invalid.
\toolName{} manages an allowlist of valid handles to avoid the necessity to keep track of a potentially endless number of invalid or revoked \keyHandle{}s.
Since the validity of a handle inherently changes over time, this information cannot be stored inside the handles.
Otherwise, an attacker could copy a valid handle and, once the original handle has been revoked, keep using the still valid copy.
Therefore, \toolName{} stores the handle allowlist in CPU-internal memory combined with its per-handle state cache entries (\handleCacheS{}).
On handle creation (\autoref{kl:sec:genHandles}), \toolName{} flags the new \handleCacheS{} entry associated with the handle as valid.
To invalidate a \keyHandle{}, \toolName{} provides a new CPU instruction that takes the handle address as input.
The instruction unwraps and checks the \keyHandle{} (cf.~\autoref{kl:sec:concept_encdec}) before flagging the respective \handleCacheS{} entry as invalid.
We discuss  \toolName{}'s current revocation strategies and who can revoke handles in \autoref{kl:appdx:revokeStrat}.
Additionally, \toolName{} supports a counter-based lifetime restriction for \keyHandle{}s, that limits the number of allowed en-/decrypt operations of a handle and revokes it when reaching zero.
That way, one-time or usage-limited handles can be implemented, e.g., as used for a licensing scheme in \autoref{kl:sec:carUC}.
As the counters must be updated on each operation, similar to the handle validity, they cannot be stored inside the handles.
Instead, each counter is stored in a handle's CPU-internal state cache entry (cf.~\autoref{kl:sec:cpu_mem}).

\smallsubsec{Restrict who uses Handles.}
\label{kl:p:binding}
Finally, \toolName{} implements restrictions that control \emph{by whom} a handle can be used.
First, users can define as subset of permitted CPU privilege levels, including user space, kernel space, or monitor mode.
\toolName{}'s instructions check the caller's privilege level directly via the respective CPU register. 
That way, handles can, for instance, be bound to the OS kernel, preventing user space attackers from using them.
Second, handles can be bound to a process or TEE context, making them usable only by that specific context, e.g., application.
\toolName{} associates a unique process or \ac{TEE} identifier (ID) with a \keyHandle{} to establish the binding.
On each handle-based operation, the binding is checked using the respective ID.

What CPU-level identifiers \toolName{} can securely use to uniquely distinguish process or TEE contexts depends on the ISA, OS, and TEE.
For instance, for our RISC-V \ac{ISA} with a Linux-based OS, the active user process can be identified using the SATP CPU register, which stores the address of the process-specific page table and its \ac{ASID}.
To allow the alternative binding of handles to \ac{TEE} instances, \toolName{} supports RISC-V TEEs based on \ac{PMP}, e.g., Keystone enclaves~\cite{keystone20}.
The \ac{PMP} memory partitions isolate TEEs from the OS and are identified via CPU-accessible PMP IDs.
Note that the Linux OS controls user processes and their IDs while monitor-mode software controls the \ac{PMP} partitions of \acp{TEE}.
Therefore, \toolName{} must trust the OS and monitor software for the respective binding type.
However, the \keyHandle{}s and all other restrictions stay secure even if the OS and monitor become compromised (cf.~threat model, \autoref{kl:subsec:threats}).
If an \ac{ISA} supports hardware-managed execution contexts, \toolName{} could adopt these for handle bindings that do not rely on trust in software components.
As some binding IDs might leak sensitive information, e.g., kernel addresses of a process, \toolName{} currently stores the ID in a handle's CPU-internal memory entry (\handleCacheS{}).
Alternatively, the ID could be stored within the handle (policy) in a masked way, e.g., encrypted or hashed.

\subsection{\handleCacheL{}}
\label{kl:sec:cpu_mem}

\toolName{} introduces a CPU-internal state cache for efficiently storing per-handle data, called \emph{\handleCacheL{} (\handleCacheS}).
Each entry of the \handleCacheS{} is associated with one valid \keyHandle{} and includes the handle's information of its stateful restriction policies, e.g., the current lifetime counter value.
Since handle lookups must be fast and \ac{CPU}-internal memory is expensive and limited, the \handleCacheS{} is designed as a small set-associative structure, similar to \ac{CPU} caches or the \ac{TLB}.
That is, the \handleCacheS{} is organized in \textit{sets} and \textit{ways} forming a table-like structure (shown in \autoref{kl:fig:hcb_overview}; Appendix).
To efficiently lookup entries in the cache, we split the bits of each \keyHandle{}'s unique \hIV{} into a cache index (selecting the set) and a tag, with the split depending on the implemented cache and IV size, e.g., \SI{6}{\bit} and \SI{90}{\bit}.
As described in \autoref{kl:sec:genHandles} and \autoref{kl:sec:ivGen}, \hIV{} is generated in a collision-free and \mbox{(pseudo-)random} way, such that statistically, the handles evenly spread over the cache sets.
The tag is stored as part of the handle entries.
On a handle-based operation, \toolName{} can use the index to select the correct cache set and then concurrently compare the tag in each way to pick the correct handle entry.
In \autoref{kl:appdx:swapping}, we discuss how to overcome the in-CPU size limitations of the \handleCacheS{} using memory swapping, enabling practically unlimited handles.

\subsection{Remote Key (Handle) Provisioning}
\label{kl:sec:keyImport}
So far, we explained how clients can protect local keys using \toolName{}.
However, many use cases involving symmetric keys require secure remote key sharing or provisioning.
For instance, consider a network IDS (NIDS) for TLS traffic as described in \autoref{kl:sec:motivationProblem}, which requires the TLS keys to be shared between the connection peers and the additional NIDS host.
Unfortunately, a system sharing keys with a remote host has no control on how they are used by that host.
A compromised NIDS host could abuse the shared keys to stealthily tamper with the TLS traffic.
In such cases, it would be beneficial if the keys would not directly leak to the NIDS host and could be restricted in usage, permitting only decrypt operations to enforce read-only traffic access even if the NIDS host gets compromised.

Therefore, \toolName{} foresees the integration of a \emph{\remoteKeyImporter{}} to enable such use cases.
The \remoteKeyImporter{} securely receives remote keys and usage policies and directly forwards them to the local \toolName{} handle unit for wrapping.
That way, untrusted local software, e.g., the NIDS, never gains direct access to the remote keys as they will become usable only as protected \keyHandle{}s with \toolName{}-enforced usage policies.

The \remoteKeyImporter{} requires the following features to enable such secure key sharing: an authentication mechanism, (confidential) isolation from the local system, and a direct interface to the \toolName{} CPU extension.
The authentication guarantees that remote peers can verify the \remoteKeyImporter{} and establish a secure channel for sharing keys (and usage policies), e.g., based on a key exchange protocol or hardware-based remote attestation.
The isolation and direct interface guarantee that the received remote keys do not leak to the local software and can be directly passed with their policies to \toolName{} for a secure \keyHandle{} transformation.
We deem multiple implementation variants possible, e.g., a full hardware extension, a lightweight CPU extension that cooperates with an external key manager such as a TPM~\cite{ElAtali2022}, or a trusted software module isolated by a CPU-provided \ac{TEE}.
Our prototype implementation is build on hardware-isolated Keystone enclaves~\cite{keystone20}, as discussed in \autoref{kl:sec:importerImpl}.
\section{Implementation}
\label{kl:sec:impl}

To demonstrate our concepts, we implemented an open-source RISC-V prototype of \toolName{}\footref{fn:proto} as a hardware extension to RocketChip~\cite{asanovic2016rocket} and the Chipyard framework~\cite{chipyard}.
We use Chipyard to instantiate a RocketCore \mbox{RISC-V} CPU equipped with a 5-stage in-order pipeline, and we integrate the Keystone project~\cite{keystone20} to enable \ac{TEE} support in the form of secure enclaves.
Our extension is implemented in the Chisel hardware description language and focuses on \keyHandle{}s for AES-GCM operations.

As shown in \autoref{kl:fig:hw_overview}, our prototype consists of a main steering unit, the \handleHWunit{} unit and \handleCacheS{} for handle management, the en-/decryption unit for handle-based operations, as well as an IV generation and AES-GCM unit.
The steering unit serves as the main control unit of the \toolName{} extension and integrates the new \keyHandle{} instructions using the RoCC interface provided by RocketCore:
\texttt{wrapkey}, \texttt{encrypt}, \texttt{decrypt}, and \texttt{revoke}.
The AES-GCM and IV units are used by the other units for the actual cryptographic operations.
To enable memory access, \toolName{} is connected to the memory interface of the CPU L1 caches using RocketChip's \textit{HellaCache} interface, allowing to fetch data from the cache or RAM.
In productive environments, the \innerWrapKey{} is securely generated by the \ac{CPU}'s \ac{TRNG} on startup, or loaded from a secure non-volatile storage.
In the current prototype, we load the key from memory instead.
The prototype does not yet implement the \handleCacheS{} swap memory (cf.~\autoref{kl:appdx:swapping}) and executes the instructions in a blocking way, halting the single-issue pipeline of the RocketChip CPU during computation---both could be improved in future versions.

\smallsubsec{AES-GCM Unit.}
\toolName{}'s current implementation builds on an AES hardware unit with support for AEAD in order to protect the \keyHandle{}s and provide handle-based crypto operations.
Our current prototype focuses on the frequently used GCM mode.
We implement the AES unit as a black box based on an open-source AES128-GCM implementation~\cite{Charles2013}.
In principle, other AES-GCM implementations can be used to achieve different area-performance tradeoffs (cf.~\autoref{kl:sec:area}).
Moreover, AES-GCM could be replaced or augmented with other (non-AES) AEAD ciphers, e.g., ChaCha20-Poly1305~\cite{rfc7539}.

The AES unit is only accessible by \toolName{}s new CPU instructions.
\toolName{} queries the AES unit for AES-GCM operations by transferring an AES key and the required data, e.g., plaintext/cipher, \ac{AAD}, or authentication tag.
On \keyHandle{} generation, the AES unit authentically encrypts the plaintext \userKey{} with the \innerWrapKey{}, and signs the usage policies as \ac{AAD}.
On handle-based crypto operations, the AES unit performs the user-requested operation (encrypt/decrypt) with the handle's decrypted \userKey{}.

\smallsubsec{Collision-free IV Generation.}
\label{kl:sec:ivGen}
\toolName{} must ensure that the IV's used for \keyHandle{} generation (\hIV{}) and encrypt-/decrypt operations (\dIV{}) are collision-free.
As AES-GCM is insecure under IV collisions, otherwise, attackers might try to exploit collisions to recover information on the plaintext of encrypted user data or on the GCM authentication key in order to spoof tags.
While the IVs must be unique, they need not be random.
Therefore, we implement a four tap \ac{LFSR} with $n=96$ and maximum cycle length~\cite{Ward2012} to sample \SI{96}{\bit} IVs.
The \ac{LFSR} is clocked every time a new IV is required, thus ensuring that collisions can occur only after $2^{96}$ handle generations or crypto operations.
Note that an attack purposely trying to overflow the \ac{LFSR} would require more than \num{2.5e+12} years when assuming \SI{1}{\nano\s} per operation (\SI{1}{\giga\hertz} clock), and is thus infeasible in practice.

\smallsubsec{CPU-Internal Registers and Memory.}
\label{kl:sec:internStorage}
\toolName{} adds new CPU-internal registers and SRAM to store its internal state information.
As our current implementation uses AES-128-GCM for the \keyHandle{} protection, \toolName{} adds a secure \SI{128}{\bit} register for the \innerWrapKey{} that is only accessible by \toolName{}.
We implement the \handleCacheS{} in CPU-\textit{}internal SRAM as a 2-way set-associative caching structure with \SI{64}{sets}.
Accordingly, we use the lower \SI{6}{\bit} of each \hIV{} as the index to select the set, and the remaining \SI{90}{\bit} as tag for selecting the way (cf.~\autoref{kl:fig:hcb_overview}, Appendix).
We store the tag in the respective \handleCacheS{} handle entries together with the \keyHandle{}'s \SI{64}{\bit} binding ID and the current \SI{8}{\bit} lifetime usage counter if enabled.
Our implementation does not store the validity flags of the \keyHandle{} allowlist mechanism in the \handleCacheS{} for efficiency reasons.
Instead, we implement a 128~bit-field register.
Each bit in the register indicates the validity of the \keyHandle{} associated with one of the $2x64$ \handleCacheS{} entries and is set/unset on handle creation and revocation accordingly.

\smallsubsec{Handle Wrapper.}
\label{kl:sec:handle_wrapper_impl}
The \handleHWunit{} unit is responsible for creating, unwrapping, and revoking \keyHandle{}s, as well as enforcing the associated restriction policies.
Accordingly, the \handleHWunit{} is involved in all new CPU instructions added by \toolName{}.

As shown in \autoref{kl:fig:handle_format}, \toolName{} currently implements \SI{512}{\bit} \keyHandle{}s that embed the encrypted \userKey{} and the associated usage policy.
As we build on AES with \SI{128}{\bit} keys, the cipher and tag length are \SI{128}{\bit} accordingly.
Similarly, we reserve up to \SI{128}{\bit} for \hIV{}, e.g., AES-GCM and ChaCha20-Poly1305 require \SI{96}{\bit} by default.
The lower \SI{128}{\bit} of a handle are used for the \keyHandle{}-embedded usage policy information.
The current calling convention of \toolName{}'s instructions passes handles via memory.
Therefore, the \keyHandle{} fields are \SI{64}{\bit}-aligned to enable faster CPU access.
Alternatively, as the \keyHandle{}s fit into four extended \SI{128}{\bit} registers, a future implementation could support a faster register-based calling convention.
The usage policy data is divided into five groups.
The fields are encoded as space-efficient bit-fields which select the AEAD cipher (\texttt{Algorithm}), permit en-/decrypt operations (\texttt{Crypt.Attr.}) or caller privilege levels (\texttt{Privileges}), or enable usage restrictions, e.g., process-binding or usage counters (\texttt{Feature Map}).
As process binding requires OS support to retrieve the target process ID (cf.~\autoref{kl:sec:usagePolicy}), we added an extra handle flag (\texttt{SelfBind}) that enables user processes to directly bind handles to their current process.
Process and enclave binding are overloaded using the \texttt{PMPMode} switch, i.e., only one can be used for a \keyHandle{} at a time.
The gray fields and reserved block (\texttt{R}) in \autoref{kl:fig:handle_format} indicate future extensions (cf.~\autoref{kl:sec:discussion}).

The \keyHandle{} creation is implemented by \toolName{}'s \texttt{wrapkey} CPU instruction.
It takes two memory references: one to a \SI{384}{\bit} \texttt{\handleGenStruct{}} struct, containing the user's AES \userKey{} (plaintext) and usage policy data, and one defining the output address of the resulting \keyHandle{}.
The policy data is similar to that included in the \keyHandle{} (\autoref{kl:fig:handle_format}), but can additionally include a \SI{64}{\bit} binding target ID (process or PMP ID) and an \SI{8}{\bit} usage counter.
The \texttt{keywrap} instruction transforms the \userKey{} into a valid \toolName{} \keyHandle{}.
First, the usage policy data (except of the binding ID and usage counter) is copied from the \texttt{\handleGenStruct{}} into the \keyHandle{}.
Afterwards, a new \hIV{} is generated using the \ac{LFSR}.
The \handleHWunit{} then derives a cache index based on the \hIV{} and uses it to add a new entry to the \handleCacheS{}, storing the tag, binding ID, and counter.
Afterwards, the \handleHWunit{} starts the authentic encryption operation by loading the required data into the AES-GCM module:
the \innerWrapKey{} as key, the \userKey{} as data, the usage policy as \ac{AAD}, and \hIV{}.
The resulting cipher and GCM tag are written into the \keyHandle{} together with the used \hIV{}, completing the handle generation.

On a handle-based operation (e.g., encrypt), the \handleHWunit{} is responsible for unwrapping the \keyHandle{} and enforcing its usage restrictions.
First, the \handleHWunit{} loads the cipher, GCM tag, \hIV{}, and AAD (usage policy) from the \keyHandle{} and inputs it into the AES-GCM unit to perform the decryption and signature verification using the \innerWrapKey{}.
Afterwards, the \handleHWunit{} looks up the \handleCacheS{} entry based on the cache index derived from \hIV{} and checks if the \keyHandle{} is valid and the usage policy restrictions are satisfied.
On success, the \handleHWunit{} can continue the requested operation, e.g., by forwarding the plaintext \userKey{} to the en-/decryption unit, or performing a handle revocation.

\toolName{}'s revocation instruction removes a \keyHandle{} from the CPU-internal allowlist.
First, the \handleHWunit{} unwraps and checks the handle.
If the operation is permitted (cf.~\autoref{kl:appdx:revokeStrat}), the \handleHWunit{} identifies the handle's bit-field entry in the allowlist register based on the position (set, way) of its \handleCacheS{} entry.
Finally, the valid bit is unset to revoke the handle, and the \handleCacheS{} entry can be reused.
If all \keyHandle{}s bound to a given process or PMP ID should be revoked, \toolName{} revokes the handles of all \handleCacheS{} entries with matching IDs.

\smallsubsec{Handle-based De-/Encryption Unit.}
\label{kl:sec:encdecUnit}
The de-/encryption unit performs handle-based crypto operations on user data, as described in \autoref{kl:sec:concept_encdec}.
\toolName{} implements two respective CPU instructions: \texttt{encrypt} and \texttt{decrypt}, which take a pointer to an \encdecStruct{} and the \keyHandle{} as inputs.
First, the \handleHWunit{} unit verifies the \keyHandle{}, and decrypts and forwards the contained \userKey{} to the de-/encryption unit.
The de-/encryption unit loads the required data (plaintext/cipher) and \ac{AAD} from the memory addresses given in the \encdecStruct{}, and---depending on the operation type---either generates a fresh \SI{96}{\bit} \dIV{} using the \ac{LFSR} for encryption or loads the user-given \dIV{} and authentication tag for decryption.
The unit forwards the information to the AES-GCM-128 unit which block-wise performs the requested encrypt/decrypt operation in place, i.e., directly writing to the input data address given in the I/O structure, for zero-copy processing.

\smallsubsec{TEE-based \remoteKeyImporter{}.}
\label{kl:sec:importerImpl}
We implemented the \remoteKeyImporter{} (\autoref{kl:sec:keyImport}) as a trusted Keystone enclave~\cite{keystone20}.
Keystone enclaves are hardware-isolated using RISC-V PMP and support remote attestation.
Thus, they enable remote services to verify the authenticity and security of the \remoteKeyImporter{} before sharing AES keys.
Remote services can establish a remote channel by performing a key exchange as part of the attestation process~\cite{sgx-ra-tls,SENG2020} to send the key and associated usage policy.
As \toolName{}'s instructions can be called from within Keystone enclaves, the \remoteKeyImporter{} enclave can directly transform the key into a protected \keyHandle{} and wipe the plaintext key from memory.
A local service can host a \remoteKeyImporter{} enclave instance and proxy the secure remote connection to the enclave, receiving the resulting \keyHandle{}(s) via shared memory.
The plaintext key is never leaked.
In \autoref{kl:sec:usecases}, we present two use cases enabled by \toolName{} and a \remoteKeyImporter{}.
\section{Security Analysis}
\label{kl:sec:security}

\smallsubsec{Hardware Attacks against \toolName{}.}
With \kl, we present a hardware security extension that is integrated into the \ac{CPU} microarchitecture.
Thus, it is important to make sure that our extension does not introduce new vulnerabilities to the microarchitecture.
The \kl extension can, in principle, be shared between multiple CPU cores. 
Since most of \kl's operations depend on the AES hardware unit, new \kl instructions must be stalled while the extension is busy. 
This separation of workloads prevents any cross-domain data leakage during the computation.
Though, due to the inherently limited hardware resources, it is feasible for attackers to observe the utilization of \kl{} by measuring the latency of \kl{} instructions.
This, however, only leaks information on whether another process is using \kl but not about the data itself\footnote{Note, that the length of the processed data may be leaked by timing. This is an inherent problem of variable-sized inputs.} or the AES keys, i.e., the \innerWrapKey{} or \userKey{}s.
Similarly, the \handleCacheL{} may leak information on which \keyHandle{}s have recently been used based on the timing of the handle verification.
Again this does not reveal information about the keys or the processed data.

While we currently focus on strong software-level attackers in our threat model (\autoref{kl:subsec:threats}), it is important to ensure that the AES unit and the critical hardware registers containing key material are sufficiently protected from physical attackers, e.g., by using masked implementations.
This challenge is shared with other hardware-based \keyHandle{} designs, and particularly relevant in settings where attackers can gain easy hardware access (e.g., IoT).
However, as \toolName{} is directly integrated into the CPU, it is non-trivial for attackers to launch physical attacks.
In contrast, in systems without \keyHandle{}s, physical attackers can directly leak keys from RAM.

\smallsubsec{Impact of System-level Attackers.}
\toolName{} assumes that all user and system software may eventually become compromised.
However, as described in our threat model (cf. \autoref{kl:subsec:threats}), local software is assumed to transform \userKey{}s into protected \keyHandle{}s before a compromise.
Therefore, even strong system-level attackers cannot leak plaintext keys or bypass the \keyHandle{} restrictions.
The \innerWrapKey{} is CPU-generated and accessible only by \toolName{}.
Therefore, only \toolName{} can unwrap \keyHandle{}s to access the \userKey{}s.
Furthermore, this implies that handles can only be used via \toolName{}'s CPU instructions, guaranteeing the enforcement of handle restrictions.
For remote key imports, \toolName{}'s \remoteKeyImporter{} ensures that remote services can securely send the plaintext keys and restriction policies to \toolName{}, without risking leakage or tampering.
The Provisioner is hardware-isolated from system-level attackers, supports authenticated E2EE communication, and directly forwards the keys to \toolName{} to transform them into protected \keyHandle{}s (cf.~\autoref{kl:sec:keyImport}).

The only type of handle restrictions affected by system-level attackers is process and \ac{TEE} binding (cf.~\autoref{kl:p:binding}).
Processes are managed by the OS kernel, and PMP-based \acp{TEE} by monitor-mode software.
Therefore, compromised kernel or monitor software can render these bindings ineffective, for example, by executing malicious code with the ID of a different process/\ac{TEE} to use a \keyHandle{} despite its binding.
Note, however, that the protection of the \keyHandle{}s and the remaining restrictions are secure even against such attackers.
In addition, if an \ac{ISA} supports hardware-managed execution contexts, \toolName{} could adopt these for handle bindings that do not rely on trust in software components.

\subsection{Challenges of Encrypt-/Decrypt-only}
\label{kl:sec:encdecissues}

The introduction of encryption-~and decryption-only \keyHandle{}s (cf. \autoref{kl:sec:restrictions}) poses some challenges from a security point of view.
Intuitively, cipher modes that encrypt data using a traditional block cipher like AES rely on the inverse block cipher function to decrypt the data.
This is, for example, the case for AES-ECB and AES-CBC.
However, when the block cipher is used in a stream cipher mode of operation, e.g., in AES-CTR or AES-OCF, the inverse function is not used.
Instead, the block cipher is used to generate a keystream that is XORed with the plaintext or ciphertext, resulting in identical encryption and decryption operations.
Therefore, the operations of these ciphers cannot be restricted, because attackers can use encrypt operations to decrypt data and vice versa, thus bypassing de-/encrypt-only restrictions.
For the same reason, the de-/encrypt-only handles of Intel Key Locker are insecure.
Similarly, AEAD ciphers like AES-GCM and ChaCha20-Poly1305 introduce new challenges for these usage restrictions.
In the following, we discuss how \toolName{} securely addresses them for AEAD ciphers.

\smallsubsec{Encryption-only.}
AEAD ciphers additionally perform data authentication, i.e., after data (\texttt{dt}) en-/decryption, they generate/verify an authentication tag over the cipher and optional additional data (\texttt{ad}).
Generally, in such cases, it \textit{does not} hold that
\begin{equation}
\label{kl:eq:gcm}
\textcolor{blue}{Enc}_{k,iv}(\textcolor{blue}{Enc}_{k,iv}(dt, ad), ad)= \textcolor{red}{Dec}_{k,iv}\textcolor{blue}{(Enc}_{k,iv}(dt, ad), ad)
\end{equation}
since the double-encryption on the left side of the equation would yield an authentication tag different from the encryption-decryption on the right side.
Therefore, one could assume that the encrypt-only restriction would  be trivially compatible with AEAD ciphers.
However, for constructions like AES-GCM or ChaCha20-Poly1305, it holds that
\begin{equation}
Enc_{k,iv}(Enc_{k,iv}(dt, ad), ad) = (dt, t'),
\end{equation}
with $t'$ being the authentication tag of the second encryption.
That is, the double encryption returns the correct plaintext and \textit{only} deviates in the authentication tag t' from the encryption-decryption on the right side of \autoref{kl:eq:gcm}.
Thus, if the attacker chooses to ignore the authentication tag, they can still decrypt the message using an encryption-only handle.
To avoid this, for each encryption operation that has the encryption-only restriction, we require \dIV{} to be generated collision-free in hardware, thus rendering it impossible for an attacker to choose colliding \dIV{} and perform the double-encryption to bypass the encrypt-only restriction.

\smallsubsec{Decryption-only.}
The decrypt instruction of \toolName{} does not return an authentication tag.
Instead, it compares the user-provided tag with the tag that is computed during the decryption, and indicates the result, i.e., the validity of the ciphertext, as a return value of the instruction.
While an attacker can use a decrypt-only handle to produce ciphertexts by \emph{decrypting} plaintexts under a chosen \dIV{}, the attacker cannot produce a valid authentication tag since the decryption never returns an authentication tag.
Thus, the attacker would need to forge the authentication tag without knowing the key, which, by design, is not feasible for AEAD ciphers.
\section{Evaluation}
In this section, we analyze \toolName{}'s hardware costs and measure its performance.
Moreover, we show how \toolName{}'s concepts solve the challenges of the three real-world use cases presented in \autoref{kl:sec:motivationProblem}.

\subsection{Area}
\label{kl:sec:area}
\begin{table}
\small
\centering
\caption{\label{kl:tab:hw_util}Area utilization of our \toolName{} prototype.}
\begin{tabular}{ l r r r} 
 \toprule
  & \textbf{LUTs} & \textbf{FFs} & \textbf{\% of LUT Overhead} \\ \midrule
 CPU w/o \toolName{} & \num{154594} & \num{134958} & - \\ 
 CPU w/ \toolName{} & \num{179118} & \num{138888}  & total: 15.9\%\\ \midrule
 AES GCM & \num{19428} & \num{1851} & 79.2\% \\
 Enc / Dec Unit & \num{1296} & \num{278} & 5.2\% \\
 Gen IV & \num{2} & \num{96} & 0\%\\
 \handleHWunit{} & \num{1418} & \num{164} & 5.7\%\\
 \handleCacheS{} & \num{848} & \num{624} & 3.5\% \\
 Mem. Access & \num{834} & \num{271} & 3.5\% \\ \bottomrule
\end{tabular}
\end{table}

\autoref{kl:tab:hw_util} shows the hardware utilization of \toolName{} on a Xilinx Alveo U250 data center accelerator card.
Overall, \toolName{} adds \SI{\approx 16}{\percent} \ac{LUT} overhead (\SI{\approx 24.5}{k}), and \SI{\approx 3}{\percent} \ac{FF} overhead (\SI{\approx 3.9}{k}) to the RocketChip \ac{CPU} core.
Note that the relative area overhead appears bigger due to the small size of the core itself (RocketChip is a single-issue in-order \ac{CPU}).
On larger processors, the relative overhead will be significantly smaller.
Furthermore, about 80\% of the total \ac{LUT} overhead introduced by \kl results from the AES-GCM unit.
If a smaller area overhead is preferred over performance, a more lightweight AES-GCM implementation can be used instead.
The remainder of \toolName{}'s hardware costs results from its handle wrapper, data en-/decryption logic, and the \handleCacheS{}.
Note that the \handleCacheS{} requires additional SRAM memory which is not shown in the table.
For the prototype, we implemented a 2-way set-associative \handleCacheS{} with 64 sets (128 entries).
Each entry holds \SI{162}{\bit} (cf.~\autoref{kl:fig:hcb_overview}), i.e., \SI{2.6}{\kilo\byte} of SRAM memory are added.

\subsection{Performance}
\label{kl:sec:perf}
In the following, we present microbenchmarks of \toolName{}'s new CPU instructions, and compare its en-/decryption performance to that of two common software libraries (OpenSSL, mbedTLS).

\begin{figure}
\centering
\includegraphics[width=.5\columnwidth]{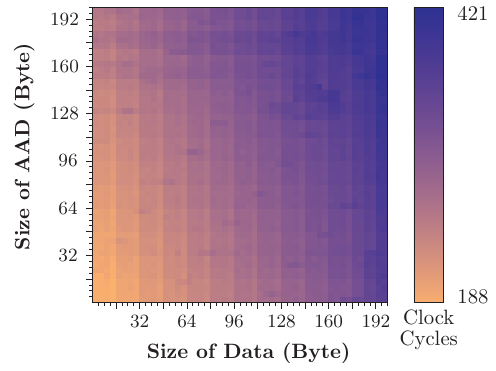}
\caption{\label{kl:fig:performance_heat}Performance of \kl encryption and decryption for varying sizes of data (x-axis) and AAD (y-axis).}
\end{figure}

\smallsubsec{Microbenchmarks.}
\autoref{kl:fig:performance_heat} shows the encryption and decryption performance of \kl with varying data and \ac{AAD} sizes in four-byte steps.
Each pixel of the heatmap is averaged over 100 iterations with random inputs.
The de-/encryption of payloads with \SI{4}{\byte} of data and \ac{AAD} takes 188 clock cycles on average, and that of payloads with \SI{200}{\byte} of data and \ac{AAD} takes 421 cycles.
Compared to an AES-GCM hardware unit without \keyHandle{} support, \toolName{}'s handle verification and unwrapping adds a static overhead of 93 clock cycles.
For a typical TLS 1.2 packet, i.e., about \SI{1500}{\byte} of data and \SI{13}{\byte} of AAD, \toolName{} requires 1439 cycles on average, i.e., \SI{\approx 6}{\percent} overhead on top of an AES-GCM unit.
Most of the latency results from the memory interactions, as the 
AES data blocks need to be loaded from and stored to memory in \SI{64}{\bit} chunks.

As shown in the figure, the latency of \kl's encryption and decryption operations increases linearly with the input sizes of data and \ac{AAD}.
However, appending further data blocks increases the latency slightly faster than adding authenticated data. 
Note that handle restrictions like process binding or decryption restrictions do not affect the latency of \kl's instructions. 
These restrictions are checked within a single clock cycle during the handle verification. 
The measurements in \autoref{kl:fig:performance_heat} show a distinct grid pattern. 
Notably, executions where the data (or AAD) length is a multiple of the AES block size (\SI{16}{\byte}) have a slightly lower latency.
This is because the AES-GCM unit and \toolName{}'s de-/encryption unit are optimized to operate on full AES blocks.
In addition, the memory access granularity and alignment can contribute to timing differences.
RocketChip's memory interface is optimized to load \SI{64}{\bit} data blocks from memory, but requires multiple accesses to load shorter blocks, e.g., for \SI{48}{\bit}, a \SI{32}{\bit} followed by \SI{16}{\bit} fetch.
A similar effect occurs for unaligned accesses as the memory unit restricts $2^w$~byte accesses to $2^w$~byte-aligned addresses.

\smallsubsec{Comparison to Software Libraries.}
To put the performance of our \kl prototype further into perspective, we compare it to software-based encryption and decryption using OpenSSL in version 1.1.1q and MBedTLS in version 2.28.2 on the RocketChip \ac{CPU}.
We used the \texttt{-O3} compiler flag to ensure that the software is optimized for high performance.
Note, however, that this is not a fair comparison, as the software implementation will always be slower than a hardware implementation.
A hardware solution such as AES-NI will perform similarly to \kl without the handle verification (cf.~\toolName{}'s static overhead in the previous section).

Like in the previous experiments, we measure the average performances over 100 runs with random input data and AAD.
For payloads with \SI{4}{\byte} of data and \ac{AAD}, the encryption has an average latency of \num{16232} clock cycles with OpenSSL and \num{22716} clock cycles with MBedTLS. 
For \SI{200}{\byte} of data and AAD, OpenSSL took \num{44372} clock cycles and MBedTLS took \num{40097} clock cycles on average.
Notably, for the software libraries, decryption is faster than encryption.
That is mostly due to the (non-TRNG) randomness initialization and the fact, that the authentication tag cannot be computed in parallel on encryptions.
For the AES-GCM decryption, we measured \num{12391} cycles and \num{9774} cycles for \SI{4}{\byte} of data and \ac{AAD}, and \num{41623} cycles and \num{37727} cycles for \SI{200}{\byte} for OpenSSL and MBedTLS respectively.

\subsection{Use Case Analysis}
\label{kl:sec:usecases}
We now revise the three real-world challenges introduced in \autoref{kl:sec:motivationProblem} and show how \toolName{} efficiently enables secure solutions for them.
For each use case, we implement a simplified proof of concept (PoC).

\smallsubsec{UC-1: Ephemeral Key-value Storage for Web Services.}
Without \toolName{}, a web server can already store encrypted user session data (e.g., credentials) in an untrusted key-value database.
While this prevents data leakage on a database compromise, without using key handles, it bares the risk that attackers can leak the key.
Existing \keyHandle{} solutions like TPMs or Intel Key Locker can protect the key to render \emph{remote} attacks ineffective, as the handle is only locally usable.
However, \emph{local} attackers can still exploit leaked handles, because the handles are valid system-wide.

With \toolName{}, the web server can securely bind the \keyHandle{} of its data encryption key to its process context.
That way, even if a \keyHandle{} is stolen by a local attacker, the CPU stops the attacker from using it, thus preventing any unauthorized data decryption.
We implemented a PoC consisting of stub web and database services, interconnected via UNIX domain sockets.
The web service generates a process-bound \keyHandle{} to authentically encrypt a value, using the associated storage key as the \ac{AAD}.
The web service later queries the stored cipher using the storage key, and decrypts and verifies it using the \keyHandle{}.
As the \keyHandle{} is only usable by the web service process and the key is part of the \ac{AAD}, the web service knows that the data is secure and correct---an optional counter can additionally prevent rollbacks.

\smallsubsec{UC-2: Automotive Feature Licensing Control.}
\label{kl:sec:carUC}
In our second use case, we considered a licensing system for automotive pay-per-use features, that must operate securely in offline mode.
The car vendors want to remotely provision lifetime-restricted signing keys to the car for authenticating feature-enable requests.
However, TPM-like solutions and Intel Key Locker lack CPU-enforced revocation, and Key Locker does not support remote key imports.
In contrast, \toolName{} can revoke key handles based on usage counters.
Moreover, \toolName{} can securely import remote keys with associated usage restrictions.
That way, vendors can provision feature keys via the \remoteKeyImporter{} (RPK), and \toolName{} enforces lifetime counters matching the licensed number of feature uses.

For our PoC, we use a simplified model consisting of a remote vendor licensing service, an automotive gateway processor running the majority of the car's software, and bus-connected computing units that control and enable features for car peripherals---here: a motor unit.
The licensing service grants the license for the sport mode while the motor unit activates the feature when requested by the gateway using a valid license key.
The gateway supports \toolName{} and a TEE-based RPK (cf. \autoref{kl:sec:importerImpl}) interacting with the licensing service via an authenticated (attested), E2EE connection~\cite{sgx-ra-tls,attestKeystone,FeIDo2022,SENG2020} to receive the license.
The license includes the number of permitted feature activations (\emph{cntr}) and a feature-specific AES key known by the motor unit, e.g., statically derived from a shared master key.
The RPK transforms the key into a \keyHandle{} and limits the handle usage to \emph{cntr} authenticated encryption (signing) operations, before wiping the key from TEE memory.
Afterwards, the handle is shared with the gateway service which can use it to request feature activation.
To activate a feature, the gateway uses the license \keyHandle{} to sign a nonce of the motor unit, and sends the request via the bus to the motor unit.
The \keyHandle{} is counter-restricted and will be revoked when the licensed number of uses has been completed.
More formally, the key derivation, handle creation, and feature-enable request for a feature \textit{ftr} can be described as:
\begin{align*}
	\textit{aes-key}_{\textit{ftr}} &:=  \textit{KDF} ( \textit{aes-key}_{\textit{master}}||\textit{name}_{\textit{ftr}} ) \\
	\textit{cntr-khndl}_{\textit{ftr}} &:=  \textit{handle-wrap}(\{cntr\}, \textit{aes-key}_{\textit{ftr}}) \\
	\textit{req}_{\textit{ftr}} &:=  \textit{auth-enc}(\{\textit{cntr-khndl}_{\textit{ftr}}\}, \textit{name}_{\textit{ftr}}||nonce)
\end{align*}
where $"||"$ is the byte string concatenation and $\textit{req}_{\textit{ftr}}$ the request.

In our PoC, our \toolName{}-enabled FPGA represents the car with a Keystone-based RPK, and the connected workstation hosts the vendor service.
The RPK enclave communicates with the vendor service via an encrypted and attested channel.
A non-TEE gateway process receives the \keyHandle{}s from the enclave and performs the feature activation protocol with the motor unit process.
We model the car bus via local UNIX domain sockets.

\smallsubsec{UC-3: Read-only TLS Traffic Monitor.}
\label{kl:sec:nids}
Finally, we envision a third-party service that offers traffic monitoring (e.g., NIDS, on-path NIPS), e.g., to find attacks in the TLS traffic of monitored workstations.
The monitoring service needs to have read-only access to the (decrypted) traffic, but should not be able to manipulate plaintexts.
Existing approaches sharing the plaintext connection keys cannot prevent such manipulations, and also TPMs or Intel Key Locker cannot securely enforce decrypt-only \keyHandle{}s.
With \toolName{}, it becomes possible to securely provision the TLS connection keys as decrypt-only handles for tamper-free traffic monitoring.

We envision the following design:
The middlebox hosting the monitoring service supports \toolName{} with an RPK.
The enterprise clients (workstations) use modified TLS libraries that securely send the AES connection keys with decrypt-only policies to the RPK.
The RPK shares the decrypt-only \keyHandle{}s with the traffic monitor, which stores them together with the associated metadata (connection info, IVs) from the clients.
The service can then look up the handles using the connection metadata, decrypt the client TLS traffic---captured on-path or forwarded by a router---and monitor the plaintext data.
If the service becomes compromised, attackers cannot tamper with the connections as \toolName{} prevents usage of the decrypt-only handles for (re-)encrypting tampered packets.

We provide a PoC for clients using the mbedTLS library and TLS 1.2 connections with AES-128-GCM ciphers.
We use our FPGA as both, the monitoring host and TLS target server.
The TLS client runs on a separate system of the same local network.
The client connects to the server to send an HTTP request.
During the TLS handshake, the TLS library establishes a secure connection to the RPK enclave running on the FPGA (cf.~previous use case) and shares the AES connection keys.
The RPK passes the key handles to the traffic monitoring process via shared memory.
For the traffic capturing, the monitoring process uses libpcap.
In our PoC, the \emph{synchronous} key sharing adds an overhead of about \SIrange{15.2}{20.4}{\percent} on top of the TLS handshake.
This is acceptable especially for mid-to-long-term connections and can be further reduced by \emph{asynchronous} sharing.
\section{Discussion}
\label{kl:sec:discussion}
In this section, we discuss \toolName{}'s portability to other platforms, as well as future use cases and extensions.

Without losing generality, \toolName{}'s current implementation focuses on RISC-V and the AES-GCM AEAD cipher.
However, \toolName{} could be transferred to other architectures and support further AEAD ciphers, e.g., ChaCha20-Poly1305 (Wireguard VPN) or \mbox{ASCON} (IoT use cases)~\cite{wireguard17,ascon21}.
When porting \toolName{} to other ISAs, the instructions must be adapted accordingly, and \toolName{}'s \mbox{RISC-V} hardware identifiers must be mapped to secure alternatives of the new ISA.
For instance, the process address space information taken from \mbox{RISC-V's} SATP register could be mapped to the CR3 and TTBRx registers of x86 and Arm.
\toolName{}'s \remoteKeyImporter{} could be implemented based on alternative platform TEEs or even as a full hardware extension, assuming it still satisfies the required isolation, authentication, and interface requirements.

\toolName{}'s \keyHandle{} format is designed to support additional restrictions.
For instance, we envision time-based lifetime restrictions that incorporate a timestamp from a trusted local clock and a lifetime duration, defining when a handle gets revoked.
Such a feature would be particularly useful to enforce periodic key renewal, e.g., when used for time-restricted authorization keys, similar to Kerberos tickets or x509 certificates.
\toolName{}'s en-/decrypt-only \keyHandle{}s could additionally be used to mimic asymmetric cryptographic properties, mapping encrypt to signing, and decrypt to verifying.
By replacing asymmetric keys with restricted \keyHandle{}s of symmetric keys, the computational workload of client devices could be decreased, especially in embedded settings.
For instance, symmetric keys could be deployed into IoT boards during manufacturing as decrypt-only \keyHandle{}s, allowing the devices to decrypt (verify) messages from the manufacturer (e.g., trusted firmware updates) while blocking message spoofing.
In principle, \toolName{} could be extended to asymmetric ciphers, protecting their private keys.
Promising candidates include the stateful post-quantum signature schemes XMSS~\cite{Huelsing2018,Buchmann2011} and LMS~\cite{McGrew2019}, since NIST recommends managing their stateful private keys securely in hardware~\cite{Cooper2020}.
\section{Related Work}
\label{fig:sec:relwork}

\toolName{} is most related to cryptographic co-processors and CPU extensions aiming at key isolation.
In addition, many projects provide \acp{TEE} or in-process isolation that can optionally protect crypto keys and their operations in separated domains.

\smallsubsec{HSMs.}
\acp{HSM} protect keys in dedicated memory and expose keys only as handles to the users.
\acp{TPM}~\cite{tpmTCG} are specific crypto co-processors (\acp{HSM}) standardized by the TCG and offered by popular vendors, such as OpenTitan~\cite{opentitan} (based on Google's Titan chip), Apple's Secure Enclave~\cite{appleSecureEnc}, and Microsoft's in-CPU Pluton~\cite{plutonAsTPM24}.
Compared to \toolName{}, these approaches provide more advanced features, e.g., measurement-based attestation, but at the cost of a more complex hardware design, sometimes including extra firmware components (e.g., OpenTitan).
\toolName{}'s lightweight CPU extension is tailored for key protection and comes at lower area costs (without extra firmware), beneficial for embedded use cases, and with an easier, more flexible software integration via its directly callable CPU instructions.
Furthermore, TPMs cannot access process or TEE identifiers as used by \toolName{} to enforce fine-grained key bindings.
In addition, except of OpenTitan, all implementations are proprietary, and \acp{HSM} like TPMs have a slower de-/encryption throughput than in-CPU accelerators like AES-NI (x86) or \toolName{}---an issue shared by virtual TPMs like vTPM~\cite{vtpm06} or fTPM~\cite{ftpmTrustZone16}.

\smallsubsec{TEE-based KMS.}
\acp{TEE} provide data and code isolation rooted in CPU extensions~\cite{sanctum16,keystone20,cure20} or dedicated co-processors~\cite{Nasahl2021}.
TEEs can be used to implement trusted applications, including key management services (KMS).
Android provides secure key storage based on Arm TrustZone~\cite{AndKeymaster} while TZ-KMS~\cite{tzkms18} uses TrustZone to implement key distribution across cloud platforms.
Similarly, Chakrabarti et~al.~\cite{badSGXkms17} designed a KMS using Intel SGX for OpenStack's Barbican while Intel developed an HSM inside SGX~\cite{sgxeHSM22} and showed with their KMRA~\cite{sgxKMRA23} how to remotely provision and use private keys of web servers inside SGX.
While these designs provide flexibility, as TEEs can run arbitrary user code, they cause performance and integration overhead, because applications must call into the TEEs for each crypto operation.
In contrast, \toolName{} supports fast, directly usable key operations via its CPU instructions.
Furthermore, TEEs have no secure notion of a caller context, i.e., they cannot enforce process or TEE binding policies like \toolName{}.
With \toolName{}, TEEs can securely protect and bind keys, such that the plaintext keys cannot be leaked and their handles are unusable outside the TEE context.
Furthermore, TEEs can be combined with \toolName{} to implement new security schemes, as shown in \autoref{kl:sec:importerImpl} and \autoref{kl:sec:usecases}.

\smallsubsec{Intel Key Locker.} \toolName{} is inspired by the proprietary Intel Key Locker CPU extension~\cite{KeyLocker2020}.
\toolName{}'s high-level design shares properties with Key Locker, but substantially extends it with concepts that overcome several fallbacks and thus enable more advanced use cases (cf.~\autoref{kl:sec:usecases} and \autoref{kl:sec:discussion}).
Key Locker supports only an AES-engine in hardware and thus has no control on how software uses its AES block operations, rendering CPU-enforced usage policies on specific AES modes or operations unfeasible.
In contrast, \toolName{} integrates full AEAD ciphers, e.g., AES-GCM, enabling for instance secure decrypt-~and encrypt-only handles as used for read-only access in \autoref{kl:sec:nids}.
Furthermore, \toolName{} provides process and TEE bindings, lifetime restrictions, as well as selective revocation, and considers the integration of a \remoteKeyImporter{} to securely import remote keys as usage-restricted \keyHandle{}s (cf.~\autoref{kl:sec:keyImport}).

\smallsubsec{Key Protection via Other Hardware Extensions.}
Prior work also aimed to protect crypto keys by leveraging more generic hardware extensions, such as memory tagging, taint tracking, or capabilities.

\textit{Memory tagging:} 
ERIM~\cite{erim19} and libmpk~\cite{libmpk19} use Intel MPK to create isolated in-process domains, e.g., to enable a web service to securely perform AES operations.
However, libmpk can be bypassed via malicious domain switches, e.g., by malicious libraries or code-reuse attacks.
ERIM requires an OS module that vets services to prevent malicious domain switches, and is bypassed by code injection attacks.
Donky~\cite{donky20} implements a similar approach for RISC-V, but replaces the OS-dependency with a new user space monitor for policy management.
However, Donky's software adoption is more complex than \toolName{} due to its domain definitions and cross-domain calls, and cannot provide hardware-protected \keyHandle{}s robust against cross-process leakage.

Taint tracking: BliMe~\cite{ElAtali2022} provides a tainting ISA extension with \ac{HSM} integration, preventing leakage of unencrypted client data.
When client data gets unencrypted by the BliMe-enabled server, BliMe's ISA extension taints the data and enforces its confidentiality.
However, software adoption of BliMe has strict requirements to avoid leakage, and the HSM is not part of BliMe's implementation.
In contrast, \toolName{} focuses on protected \keyHandle{}s with CPU-enforced usage policies and high performance crypto operations.

\textit{Capabilities:}
Memory capability systems like CHERI~\cite{cheriComparts15} or Capstone~\cite{capstone23} provide fine-grained memory isolation support, which could be used to isolate crypto keys.
However, they lack a key handle abstraction with tailored usage policies and easy integration into existing real-world applications.
Capacity~\cite{capacity23} implements object capabilities based on Arm PA, Arm MTE, and a kernel extension for fine-grained protection of file-based resources and memory.
While Capacity can isolate a private key file and its memory buffers, Capacity relies on the security of the OS for \emph{all} its policies and does not enforce crypto-specific policies, e.g., decrypt-only handles.
\section{Conclusion}
The protection of cryptographic keys is essential to the security of higher-level security schemes.
Therefore, several existing designs remove the plaintext keys from unprotected memory to prevent leakage.
Instead, they replace the keys with software-usable \keyHandle{}s that hide the plaintext from users and attackers.
However, existing approaches are limited in settings requiring high-performance and fine-grained control on the usage, revocation, and deployment of \keyHandle{}s.
For instance, external devices like TPMs feature a rather slow performance and limited insights into the CPU's execution context, while CPU extensions like Intel KeyLocker lack control on how and by whom handles are used and might not support remote keys.
In this paper, we therefore introduced \toolName{}, a lightweight but high-performance CPU extension for protected \keyHandle{}s with CPU-enforced usage and revocation control.
\toolName{} combines CPU-exposed context information with new per-handle custom state to control \emph{how} and \emph{by whom} handles can be used, and when each handle is revoked.
Furthermore, \toolName{} enables advanced remote use cases by supporting a trusted key provisioner that securely transforms remote keys to local \keyHandle{}s.
Our open-source RISC-V prototype demonstrates how \toolName{} enables new security schemes ranging from high-performance networking use cases (e.g., TLS traffic monitoring) to embedded feature licensing schemes.

\appendix

\section{Key Handle Revocation Strategies}
\label{kl:appdx:revokeStrat}
\toolName{} enables flexible revocation strategies, depending on the usage policies of the respective \keyHandle{}s.
Except for TEE-bound handles, software with higher CPU privilege levels is permitted to revoke a handle, even if not included in the usage policy.
For instance, an OS kernel should be able to clean up the handles of a terminating process, though it might not be permitted to use them directly, similar to how SMAP works for user memory pages~\cite{lwnSMAP}.
\toolName{} currently enforces the following revocation policy:

For unbound \keyHandle{}s, i.e., not bound to an execution context (cf.~\autoref{kl:p:binding}), \toolName{} permits handle revocation by any software with a CPU privilege level $\geq$ the smallest level permitted by the handle's usage policy.
As unbound handles are meant to be easily shared across processes via memory, a flexible revocation strategy is reasonable.
Note that a \keyHandle{} cannot be easily guessed by a local attacker process, e.g., to use or revoke an unbound handle, because of the unpredictable \hIV{}, cipher, and tag data included in the handle.
For process-bound handles, \toolName{} permits revocation by (1.)~the process to which it is bound, and (2.)~every context with privileges $>$ the smallest level permitted by the handle policy.
\toolName{} provides a CPU instruction for revoking all handles of a given process ID, enabling, e.g., an OS on a process termination to clean up all handles bound to that process, without the need to track the handles in memory.
For handles bound to a PMP-based TEE (e.g., Keystone enclave), \toolName{} stays in line with the TEE threat model by only permitting revocation by (1.)~the respective PMP context (TEE), and (2.)~the PMP-managing monitor mode software (for TEE cleanup).
In particular, the untrusted OS is not permitted to revoke handles bound to an isolated TEE.

System software integration of \toolName{}'s revocation mechanisms can include extensions to the OS and monitor software.
The process and TEE termination handlers can be extended to revoke \toolName{}'s PID/PMP-based revocation instruction to invalidate all \keyHandle{}'s bound to the terminating process or TEE.
Unbound \keyHandle{}s can be cleaned up by the users, or by the OS and monitor software.
The OS and monitor software could create unbound handles on behalf of a user application/TEE, such that they can keep track of the respective \keyHandle{}s in descriptor tables, similar to those used for file or socket descriptors.
Alternatively, \toolName{} could internally store the process/TEE IDs of the \keyHandle{}s' creator contexts, and revoke them together with the bound handles.
We skip the engineering details as invoking \toolName{}'s revocation instructions is straightforward and the OS/monitor integrations are not specific to \toolName{} and do not require new concepts.

\begin{figure}
	\centering
	\includegraphics[width=.65\columnwidth]{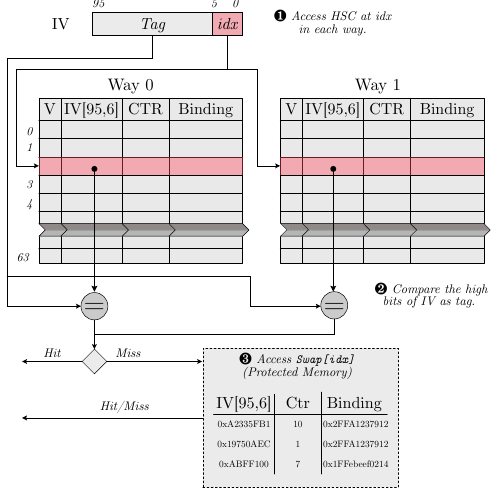}
	\caption{\label{kl:fig:hcb_overview}Overview of \toolName{}'s \handleCacheL{} and its IV-based indexing, assuming 96~bit handle IVs (\texorpdfstring{IV\textsubscript{handle}}{IV handle}).}
\end{figure}

\section{\handleCacheS{} Swapping to RAM (optional)}
\label{kl:appdx:swapping}
In principle, with \keyHandle{}s protected using AES-GCM, \toolName{}'s \handleCacheL{} (\handleCacheS{}) could allow for up to $2^{96}$ valid \keyHandle{}s, as it derives the cache indices and tags from the \SI{96}{\bit} GCM handle IVs (cf.~\autoref{kl:fig:hcb_overview}).
However, since on-chip area for CPU-internal memory is expensive, a much smaller \handleCacheS{} is preferred, e.g., a 2-way cache with \SI{64}{sets} as used in our prototype (cf.~\autoref{kl:sec:internStorage}).
If the cache is full, no more handles can be created until the next revocation.
To still allow for a practically unlimited number of handles, the \handleCacheS{} can optionally be extended to act as an LRU (least recently used) cache that swaps handles to memory when the \handleCacheS{} set is occupied.
For this, a memory region must be reserved as a handle swap region.
When entries are swapped out to non-secure memory, they are authentically encrypted using a CPU-internal storage key statically derived from the \innerWrapKey{}, with the \hIV{} tag and index being signed as \ac{AAD}.
In addition, a monotonic counter value must be included in the AAD to prevent rollback attacks against the swap region~\cite{rote17}.
With swapping enabled, \toolName{} must inspect the swap memory on a cache miss using \hIV{} and swap in decrypted entries on demand.
As LRU-based swapping is well-known (e.g., page tables) and not specific to \toolName{}, we skip further details of a potential implementation.
\bibliographystyle{alpha}
\bibliography{main}

\newcommand{\etalchar}[1]{$^{#1}$}
\begin{thebibliography}{GAMG{\etalchar{+}}23}

\bibitem[AAB{\etalchar{+}}16]{asanovic2016rocket}
Krste Asanovic, Rimas Avizienis, Jonathan Bachrach, Scott Beamer, David
  Biancolin, Christopher Celio, Henry Cook, Daniel Dabbelt, John Hauser, Adam
  Izraelevitz, et~al.
\newblock {The rocket chip generator}.
\newblock {\em EECS Department, University of California, Berkeley, Tech. Rep.
  UCB/EECS-2016-17}, 4, 2016.

\bibitem[ABG{\etalchar{+}}20]{chipyard}
Alon Amid, David Biancolin, Abraham Gonzalez, Daniel Grubb, Sagar Karandikar,
  Harrison Liew, Albert Magyar, Howard Mao, Albert Ou, Nathan Pemberton, Paul
  Rigge, Colin Schmidt, John Wright, Jerry Zhao, Yakun~Sophia Shao, Krste
  Asanovi\'{c}, and Borivoje Nikoli\'{c}.
\newblock {Chipyard: Integrated Design, Simulation, and Implementation
  Framework for Custom SoCs}.
\newblock {\em IEEE Micro}, 40(4), 2020.

\bibitem[{App}24]{appleSecureEnc}
{Apple Inc.}
\newblock {Protecting keys with the Secure Enclave}.
\newblock
  \url{https://developer.apple.com/documentation/security/certificate\_key\_and\_trust\_services/keys/protecting\_keys\_with\_the\_secure\_enclave},
  2024.

\bibitem[BBD{\etalchar{+}}21]{cure20}
Raad Bahmani, Ferdinand Brasser, Ghada Dessouky, Patrick Jauernig, Matthias
  Klimmek, Ahmad-Reza Sadeghi, and Emmanuel Stapf.
\newblock {CURE: A Security Architecture with CUstomizable and Resilient
  Enclaves}.
\newblock In {\em USENIX Security}, 2021.

\bibitem[BCG{\etalchar{+}}06]{vtpm06}
Stefan Berger, Ramon Caceres, Kenneth~A. Goldman, Ronald Perez, Reiner Sailer,
  and Leendert van Doorn.
\newblock {vTPM: Virtualizing the Trusted Platform Module}.
\newblock In {\em USENIX Security}, 2006.

\bibitem[BDH11]{Buchmann2011}
Johannes Buchmann, Erik Dahmen, and Andreas H{\"u}lsing.
\newblock {XMSS-a practical forward secure signature scheme based on minimal
  security assumptions}.
\newblock In {\em {Post-Quantum Cryptography: 4th International Workshop,
  PQCrypto 2011, Taipei, Taiwan, November 29--December 2, 2011. Proceedings
  4}}, pages 117--129. Springer, 2011.

\bibitem[Ber23]{Charles2013}
Luca Berghella.
\newblock {AES-GCM 128-192-256 bits}.
\newblock \url{https://github.com/BLu85/AES-GCM-128-192-256-bits}, March 2023.
\newblock Commit: 0b9bee5.

\bibitem[BJKS21]{Buhren2021}
Robert Buhren, Hans~Niklas Jacob, Thilo Krachenfels, and Jean{-}Pierre Seifert.
\newblock {One Glitch to Rule Them All: Fault Injection Attacks Against AMD's
  Secure Encrypted Virtualization}.
\newblock In Yongdae Kim, Jong Kim, Giovanni Vigna, and Elaine Shi, editors,
  {\em {CCS} '21: 2021 {ACM} {SIGSAC} Conference on Computer and Communications
  Security, Virtual Event, Republic of Korea, November 15 - 19, 2021}, pages
  2875--2889. {ACM}, 2021.

\bibitem[BMW{\etalchar{+}}18]{Bulck2018}
Jo~Van Bulck, Marina Minkin, Ofir Weisse, Daniel Genkin, Baris Kasikci, Frank
  Piessens, Mark Silberstein, Thomas~F. Wenisch, Yuval Yarom, and Raoul
  Strackx.
\newblock {Foreshadow: Extracting the Keys to the Intel SGX Kingdom with
  Transient Out-of-Order Execution}.
\newblock In {\em USENIX Security}, 2018.

\bibitem[CAD{\etalchar{+}}20]{Cooper2020}
David~A Cooper, Daniel~C Apon, Quynh~H Dang, Michael~S Davidson, Morris~J
  Dworkin, Carl~A Miller, et~al.
\newblock {Recommendation for stateful hash-based signature schemes}.
\newblock {\em {NIST Special Publication}}, 800(208):800--208, 2020.

\bibitem[CBV17]{badSGXkms17}
Somnath Chakrabarti, Brandon Baker, and Mona Vij.
\newblock {Intel SGX Enabled Key Manager Service with OpenStack Barbican},
  2017.

\bibitem[CCX{\etalchar{+}}20]{Chen2020}
Guoxing Chen, Sanchuan Chen, Yuan Xiao, Yinqian Zhang, Zhiqiang Lin, and
  Ten{-}Hwang Lai.
\newblock {SgxPectre: Stealing Intel Secrets From {SGX} Enclaves via
  Speculative Execution}.
\newblock {\em {IEEE} Secur. Priv.}, 18(3), 2020.

\bibitem[CHB19]{chakraborty2019simtpm}
Dhiman Chakraborty, Lucjan Hanzlik, and Sven Bugiel.
\newblock {simTPM: User-centric TPM for Mobile Devices}.
\newblock In {\em 28th {USENIX} Security Symposium}, 2019.

\bibitem[{Chr}21]{ascon21}
{Christoph Dobraunig and Maria Eichlseder and Florian Mendel and Martin
  Schl\"{a}ffer}.
\newblock {Ascon v1.2: Lightweight Authenticated Encryption and Hashing}.
\newblock {\em {J. Cryptol.}}, 34(3), 2021.

\bibitem[CLD16]{sanctum16}
Victor Costan, Ilia Lebedev, and Srinivas Devadas.
\newblock {Sanctum: Minimal Hardware Extensions for Strong Software Isolation}.
\newblock In {\em USENIX Security}, 2016.

\bibitem[Cor12]{lwnSMAP}
Jonathan Corbet.
\newblock {Supervisor mode access prevention}.
\newblock \url{https://lwn.net/Articles/517475/}, 2012.

\bibitem[DDCNL23]{capacity23}
Kha Dinh~Duy, Kyuwon Cho, Taehyun Noh, and Hojoon Lee.
\newblock {Capacity: Cryptographically-Enforced In-Process Capabilities for
  Modern ARM Architectures}.
\newblock In {\em ACM SIGSAC Conference on Computer and Communications
  Security}, 2023.

\bibitem[Don17]{wireguard17}
Jason~A. Donenfeld.
\newblock {WireGuard: Next Generation Kernel Network Tunnel}.
\newblock In {\em Network and Distributed System Security Symposium ({NDSS})},
  2017.

\bibitem[DWY{\etalchar{+}}19]{lightbox19}
Huayi Duan, Cong Wang, Xingliang Yuan, Yajin Zhou, Qian Wang, and Kui Ren.
\newblock {LightBox: Full-Stack Protected Stateful Middlebox at Lightning
  Speed}.
\newblock In {\em Proceedings of the ACM SIGSAC Conference on Computer and
  Communications Security}, 2019.

\bibitem[EGLA22]{ElAtali2022}
Hossam ElAtali, Lachlan~J. Gunn, Hans Liljestrand, and N.~Asokan.
\newblock {BliMe: Verifiably Secure Outsourced Computation with
  Hardware-Enforced Taint Tracking}.
\newblock {\em CoRR}, abs/2204.09649, 2022.

\bibitem[Enc21]{attestKeystone}
Keystone Enclave.
\newblock {Attestation -- Keystone Enclave}.
\newblock
  \url{https://docs.keystone-enclave.org/en/latest/Keystone-Applications/Attestation.html},
  2021.

\bibitem[GAMG{\etalchar{+}}23]{compasec23}
Johannes Geier, Lukas Auer, Daniel Mueller-Gritschneder, Uzair Sharif, and Ulf
  Schlichtmann.
\newblock {CompaSeC: A Compiler-Assisted Security Countermeasure to Address
  Instruction Skip Fault Attacks on RISC-V}.
\newblock In {\em Asia and South Pacific Design Automation Conference}. ACM,
  2023.

\bibitem[GESM17]{Goetzfried2017}
Johannes G{\"{o}}tzfried, Moritz Eckert, Sebastian Schinzel, and Tilo
  M{\"{u}}ller.
\newblock {Cache Attacks on Intel SGX}.
\newblock In Cristiano Giuffrida and Angelos Stavrou, editors, {\em European
  Workshop on Systems Security, {EUROSEC}}. {ACM}, 2017.

\bibitem[{Goo}24]{AndKeymaster}
{Google LLC}.
\newblock {Hardware-backed Keystore | Android Open Source Project}.
\newblock \url{https://source.android.com/security/keystore}, 2024.

\bibitem[Gro19]{tpmTCG}
Trusted~Computing Group.
\newblock {Trusted Platform Module (TPM)}.
\newblock
  \url{https://trustedcomputinggroup.org/work-groups/trusted-platform-module/},
  2019.

\bibitem[HBG{\etalchar{+}}18]{Huelsing2018}
Andreas H{\"{u}}lsing, Denis Butin, Stefan{-}Lukas Gazdag, Joost Rijneveld, and
  Aziz Mohaisen.
\newblock {XMSS: eXtended Merkle Signature Scheme}.
\newblock {\em {RFC}}, 8391, 2018.

\bibitem[{Int}]{sgxeHSM22}
{Intel Corporation}.
\newblock {eHSM (SGX Enclave Based Hardware Security Module)}.
\newblock \url{https://github.com/intel/ehsm}.

\bibitem[{Int}23]{sgxKMRA23}
{Intel Corporation}.
\newblock {Intel Software Guard Extensions (Intel SGX) – Key Management
  Reference Application (KMRA)}.
\newblock 2023.
\newblock
  \url{https://networkbuilders.intel.com/solutionslibrary/intel-sgx-kmra-on-intel-xeon-processors-technology-guide}.

\bibitem[{Jin}05]{JWCP2005}
{Jinpeng Wei and Calton Pu}.
\newblock {TOCTTOU Vulnerabilities in UNIX-Style File Systems: An Anatomical
  Study}.
\newblock In Garth Gibson, editor, {\em {Proceedings of the FAST '05 Conference
  on File and Storage Technologies, December 13-16, 2005, San Francisco,
  California, USA}}. {USENIX}, 2005.

\bibitem[Key20]{KeyLocker2020}
{\em {Intel Key Locker Specification}}, 343965-001us, rev. 1.0 edition, 2020.
\newblock
  \url{https://www.intel.com/content/www/us/en/develop/download/intel-key-locker-specification.html}.

\bibitem[KSC{\etalchar{+}}18]{sgx-ra-tls}
Thomas Knauth, Michael Steiner, Somnath Chakrabarti, Li~Lei, Cedric Xing, and
  Mona Vij.
\newblock {Integrating Remote Attestation with Transport Layer Security}.
\newblock {\em CoRR}, abs/1801.05863, 2018.

\bibitem[LHX18]{tzkms18}
Shiyu Luo, Zhichao Hua, and Yubin Xia.
\newblock {TZ-KMS: A Secure Key Management Service for Joint Cloud Computing
  with ARM TrustZone}.
\newblock In {\em IEEE Symposium on Service-Oriented System Engineering}, 2018.

\bibitem[LKS{\etalchar{+}}20]{keystone20}
Dayeol Lee, David Kohlbrenner, Shweta Shinde, Krste Asanovic, and Dawn Song.
\newblock {Keystone: An Open Framework for Architecting Trusted Execution
  Environments}.
\newblock In {\em European Conference on Computer Systems}, EuroSys, 2020.

\bibitem[LSL{\etalchar{+}}19]{lee2019matls}
Hyunwoo Lee, Zach Smith, Junghwan Lim, Gyeongjae Choi, Selin Chun, Taejoong
  Chung, and Ted~Taekyoung Kwon.
\newblock {maTLS: How to Make TLS middlebox-aware?}
\newblock In {\em 26th Annual Network and Distributed System Security
  Symposium}, 2019.

\bibitem[MAK{\etalchar{+}}17]{rote17}
Sinisa Matetic, Mansoor Ahmed, Kari Kostiainen, Aritra Dhar, David Sommer,
  Arthur Gervais, Ari Juels, and Srdjan Capkun.
\newblock {ROTE: Rollback Protection for Trusted Execution}.
\newblock In {\em {USENIX} Security Symposium ({USENIX} Security 17)}, 2017.

\bibitem[MCF19]{McGrew2019}
David McGrew, Michael Curcio, and Scott Fluhrer.
\newblock {RFC 8554: Leighton-Micali hash-based signatures}, 2019.

\bibitem[{Mic}22]{tpmboot}
{Microsoft Corporation}.
\newblock {Measured boot and host attestation}.
\newblock
  \url{https://learn.microsoft.com/en-us/azure/security/fundamentals/measured-boot-host-attestation},
  2022.

\bibitem[{Mic}24]{plutonAsTPM24}
{Microsoft Corporation}.
\newblock {Microsoft Pluton as Trusted Platform Module}.
\newblock
  \url{https://learn.microsoft.com/en-us/windows/security/hardware-security/pluton/pluton-as-tpm},
  2024.

\bibitem[NL15]{rfc7539}
Yoav Nir and Adam Langley.
\newblock {ChaCha20 and Poly1305 for IETF Protocols}.
\newblock \url{https://www.rfc-editor.org/info/rfc7539}, 2015.
\newblock RFC 7539.

\bibitem[NSUH21]{pmpFaultinj20}
Shoei Nashimoto, Daisuke Suzuki, Rei Ueno, and Naofumi Homma.
\newblock {Bypassing Isolated Execution on RISC-V using Side-Channel-Assisted
  Fault-Injection and Its Countermeasure}.
\newblock {\em IACR Transactions on Cryptographic Hardware and Embedded
  Systems}, 2022(1), 2021.

\bibitem[NSWM21]{Nasahl2021}
Pascal Nasahl, Robert Schilling, Mario Werner, and Stefan Mangard.
\newblock {HECTOR-V: A Heterogeneous CPU Architecture for a Secure RISC-V
  Execution Environment}.
\newblock In {\em {ACM Asia Conference on Computer and Communications
  Security}}, 2021.

\bibitem[ope]{opentitan}
{Open source silicon root of trust (RoT) | OpenTitan}.
\newblock \url{https://opentitan.org/}.

\bibitem[PLPR18]{safebricks18}
Rishabh Poddar, Chang Lan, Raluca~Ada Popa, and Sylvia Ratnasamy.
\newblock {SafeBricks: Shielding Network Functions in the Cloud}.
\newblock In {\em 15th USENIX Symposium on Networked Systems Design and
  Implementation}, 2018.

\bibitem[PLX{\etalchar{+}}19]{libmpk19}
Soyeon Park, Sangho Lee, Wen Xu, HyunGon Moon, and Taesoo Kim.
\newblock {libmpk: Software Abstraction for Intel Memory Protection Keys (Intel
  MPK)}.
\newblock In {\em 2019 USENIX Annual Technical Conference (USENIX ATC 19)},
  2019.

\bibitem[RSW{\etalchar{+}}16]{ftpmTrustZone16}
Himanshu Raj, Stefan Saroiu, Alec Wolman, Ronald Aigner, Jeremiah Cox, Paul
  England, Chris Fenner, Kinshuman Kinshumann, Jork Loeser, Dennis Mattoon,
  Magnus Nystrom, David Robinson, Rob Spiger, Stefan Thom, and David Wooten.
\newblock {fTPM: A Software-Only Implementation of a TPM Chip}.
\newblock In {\em USENIX Security}, 2016.

\bibitem[SDH{\etalchar{+}}22]{FeIDo2022}
Fabian Schwarz, Khue Do, Gunnar Heide, Lucjan Hanzlik, and Christian Rossow.
\newblock {FeIDo: Recoverable FIDO2 Tokens Using Electronic IDs}.
\newblock In {\em Proceedings of the ACM SIGSAC Conference on Computer and
  Communications Security}, 2022.

\bibitem[SFSG23]{Stolz2023}
Florian Stolz, Marc Fyrbiak, Pascal Sasdrich, and Tim G{\"{u}}neysu.
\newblock {Recommendation for a Holistic Secure Embedded {ISA} Extension}.
\newblock In Mehdi Tibouchi and Xiaofeng Wang, editors, {\em Applied
  Cryptography and Network Security - 21st International Conference, {ACNS}
  2023, Kyoto, Japan, June 19-22, 2023, Proceedings, Part {II}}, volume 13906
  of {\em Lecture Notes in Computer Science}, pages 62--84. Springer, 2023.

\bibitem[SR20]{SENG2020}
Fabian Schwarz and Christian Rossow.
\newblock {SENG, the SGX-Enforcing Network Gateway: Authorizing Communication
  from Shielded Clients}.
\newblock In {\em 29th {USENIX} Security Symposium}, 2020.

\bibitem[SWS{\etalchar{+}}20]{donky20}
David Schrammel, Samuel Weiser, Stefan Steinegger, Martin Schwarzl, Michael
  Schwarz, Stefan Mangard, and Daniel Gruss.
\newblock {Donky: Domain Keys {\textendash} Efficient In-Process Isolation for
  RISC-V and x86}.
\newblock In {\em USENIX Security}, 2020.

\bibitem[{Syn}20]{heartbleed14}
{Synopsys, Inc.}
\newblock {Heartbleed Bug}.
\newblock \url{https://heartbleed.com/}, 2020.

\bibitem[TBE{\etalchar{+}}21]{Trouchkine2021}
Thomas Trouchkine, S{\'{e}}banjila~Kevin Bukasa, Mathieu Escouteloup, Ronan
  Lashermes, and Guillaume Bouffard.
\newblock {Electromagnetic fault injection against a complex CPU, toward new
  micro-architectural fault models}.
\newblock {\em J. Cryptogr. Eng.}, 11(4):353--367, 2021.

\bibitem[VOED{\etalchar{+}}19]{erim19}
Anjo Vahldiek-Oberwagner, Eslam Elnikety, Nuno~O. Duarte, Michael Sammler,
  Peter Druschel, and Deepak Garg.
\newblock {ERIM: Secure, Efficient In-process Isolation with Protection Keys
  (MPK)}.
\newblock In {\em USENIX Security}, 2019.

\bibitem[WM12]{Ward2012}
R.~W. Ward and T.C.A. Molteno.
\newblock {Table of Linear Feedback Shift Registers}.
\newblock Technical Report 2012-1, Department of Physics, University of Otago,
  2012.

\bibitem[WWN{\etalchar{+}}15]{cheriComparts15}
Robert~N.M. Watson, Jonathan Woodruff, Peter~G. Neumann, Simon~W. Moore,
  Jonathan Anderson, David Chisnall, Nirav Dave, Brooks Davis, Khilan Gudka,
  Ben Laurie, Steven~J. Murdoch, Robert Norton, Michael Roe, Stacey Son, and
  Munraj Vadera.
\newblock {CHERI: A Hybrid Capability-System Architecture for Scalable Software
  Compartmentalization}.
\newblock In {\em IEEE Symposium on Security and Privacy}, 2015.

\bibitem[YWB{\etalchar{+}}23]{capstone23}
Jason~Zhijingcheng Yu, Conrad Watt, Aditya Badole, Trevor~E. Carlson, and
  Prateek Saxena.
\newblock {Capstone: A Capability-based Foundation for Trustless Secure Memory
  Access}.
\newblock In {\em USENIX Security}, 2023.

\end{thebibliography}


\end{document}